%% file: main.tex
\documentclass[twocolumn,tighten]{aastex63}
\usepackage[T1]{fontenc}
\usepackage{enumitem}
\usepackage{xspace}
\usepackage[multiple]{footmisc}
\usepackage{framed}

\setlist{noitemsep} 

\NewPageAfterKeywords

\def\m87{M87$^*$\xspace}
\def\sgra{Sgr~A$^*$\xspace}
\def\comrade{\texttt{Comrade.jl}\xspace}

\def\lsim{\mathrel{\raise.3ex\hbox{$<$\kern-.75em\lower1ex\hbox{$\sim$}}}}
\def\gsim{\mathrel{\raise.3ex\hbox{$>$\kern-.75em\lower1ex\hbox{$\sim$}}}}

\defcitealias{EHTC_M87_I}{\m87~I}
\defcitealias{EHTC_M87_II}{\m87~II}
\defcitealias{EHTC_M87_III}{\m87~III}
\defcitealias{EHTC_M87_IV}{\m87~IV}
\defcitealias{EHTC_M87_V}{\m87~V}
\defcitealias{EHTC_M87_VI}{\m87~VI}
\defcitealias{EHTC_M87_VII}{\m87~VII}
\defcitealias{EHTC_M87_VIII}{\m87~VIII}
\defcitealias{EHTC_SgrA_I}{\sgra~I}
\defcitealias{EHTC_SgrA_II}{\sgra~II}
\defcitealias{EHTC_SgrA_III}{\sgra~III}
\defcitealias{EHTC_SgrA_IV}{\sgra~IV}
\defcitealias{EHTC_SgrA_V}{\sgra~V}
\defcitealias{EHTC_SgrA_VI}{\sgra~VI}

\defcitealias{Issaoun_2022}{Issaoun, Wielgus et al. 2022}

\defcitealias{HPC}{HPC White Paper}

\begin{document}

\title{Key Science Goals for the Next-Generation Event Horizon Telescope}
\correspondingauthor{Michael~D.~Johnson}
\email{mjohnson@cfa.harvard.edu}

\author[0000-0002-4120-3029]{Michael D. Johnson}
\affiliation{Center for Astrophysics $|$ Harvard \& Smithsonian, 60 Garden Street, Cambridge, MA 02138, USA}
\affiliation{Black Hole Initiative at Harvard University, 20 Garden Street, Cambridge, MA 02138, USA}

%%%%%%%%%%%%%%%%%%%%%%%%%%%%%%%%%%%%
% SWG Coordinators
%%%%%%%%%%%%%%%%%%%%%%%%%%%%%%%%%%%%
\author[0000-0002-9475-4254]{Kazunori Akiyama}
\affiliation{Massachusetts Institute of Technology Haystack Observatory, 99 Millstone Road, Westford, MA 01886, USA}
\affiliation{National Astronomical Observatory of Japan, 2-21-1 Osawa, Mitaka, Tokyo 181-8588, Japan}
\affiliation{Black Hole Initiative at Harvard University, 20 Garden Street, Cambridge, MA 02138, USA}

\author[0000-0002-9030-642X]{Lindy~Blackburn}
\affiliation{Center for Astrophysics $|$ Harvard \& Smithsonian, 60 Garden Street, Cambridge, MA 02138, USA}
\affiliation{Black Hole Initiative at Harvard University, 20 Garden Street, Cambridge, MA 02138, USA}

\author[0000-0003-0077-4367]{Katherine L. Bouman}
\affiliation{California Institute of Technology, 1200 East California Boulevard, Pasadena, CA 91125, USA}

\author[0000-0002-3351-760X]{Avery E. Broderick}
\affiliation{Perimeter Institute for Theoretical Physics, 31 Caroline Street North, Waterloo, ON, N2L 2Y5, Canada}
\affiliation{Department of Physics and Astronomy, University of Waterloo, 200 University Avenue West, Waterloo, ON, N2L 3G1, Canada}
\affiliation{Waterloo Centre for Astrophysics, University of Waterloo, Waterloo, ON, N2L 3G1, Canada}

\author[0000-0003-0553-0433]{Vitor Cardoso}
\affiliation{Niels Bohr International Academy, Niels Bohr Institute, Blegdamsvej 17, 2100 Copenhagen, Denmark}
\affiliation{CENTRA, Departamento de F\'{\i}sica, Instituto Superior T\'ecnico -- IST, Universidade de Lisboa -- UL,
Avenida Rovisco Pais 1, 1049 Lisboa, Portugal}

\author[0000-0002-5654-2744]{R.~P.~Fender}
\affiliation{Astrophysics, Department of Physics, University of Oxford, Keble Road, Oxford, OX1 3RH, UK \\}
\affiliation{Department of Astronomy, University of Cape Town, Private Bag X3, Rondebosch 7701, South Africa \\}

\author[0000-0002-1827-1656]{Christian M. Fromm}
\affiliation{Institut f\"ur Theoretische Physik und Astrophysik, Universit\"at W\"urzburg, Emil-Fischer-Strasse 31, 97074
W\"urzburg, Germany}
\affiliation{Institut f\"ur Theoretische Physik, Goethe Universit\"at, Max-von-Laue-Str. 1, D-60438 Frankfurt, Germany}
\affiliation{Max-Planck-Institut f\"ur Radioastronomie, Auf dem H\"ugel 69, D-53121 Bonn, Germany}

\author[0000-0002-6429-3872]{Peter Galison}
\affiliation{Black Hole Initiative at Harvard University, 20 Garden Street, Cambridge, MA 02138, USA}
\affiliation{Department of History of Science, Harvard University, Cambridge, MA 02138, USA}
\affiliation{Department of Physics, Harvard University, Cambridge, MA 02138, USA}

\author[0000-0003-4190-7613]{Jos\'e L. G\'omez}
\affiliation{Instituto de Astrof\'{\i}sica de Andaluc\'{\i}a-CSIC, Glorieta de la Astronom\'{\i}a s/n, E-18008 Granada, Spain}

\author[0000-0001-6803-2138]{Daryl Haggard}
\affiliation{Department of Physics, McGill University, 3600 rue University, Montréal, QC H3A 2T8, Canada}
\affiliation{Trottier Space Institute at McGill, 3550 rue University, Montréal, QC H3A 2A7, Canada}

\author[0000-0003-1315-3412]{Matthew L. Lister}
\affiliation{Department of Physics and Astronomy, Purdue University, 525 Northwestern Avenue, West Lafayette, IN 47907, USA}

\author[0000-0003-1622-1484]{Andrei P. Lobanov}
\affiliation{Max-Planck-Institut f\"ur Radioastronomie, Auf dem H\"ugel 69, D-53121 Bonn, Germany}

\author[0000-0001-9564-0876]{Sera Markoff}
\affiliation{Anton Pannekoek Institute for Astronomy, University of Amsterdam, Science Park 904, 1098 XH, Amsterdam, The Netherlands}
\affiliation{Gravitation and Astroparticle Physics Amsterdam (GRAPPA) Institute, University of Amsterdam, Science Park 904, 1098 XH Amsterdam, The Netherlands}

\author[0000-0002-1919-2730]{Ramesh Narayan}
\affiliation{Center for Astrophysics $|$ Harvard \& Smithsonian, 60 Garden Street, Cambridge, MA 02138, USA}
\affiliation{Black Hole Initiative at Harvard University, 20 Garden Street, Cambridge, MA 02138, USA}

\author[0000-0002-5554-8896]{Priyamvada~Natarajan}
\affiliation{Department of Astronomy, Yale University, 52 Hillhouse Avenue, New Haven, CT 06511, USA}
\affiliation{Department of Physics, Yale University, P.O. Box 208121, New Haven, CT 06520, USA}
\affiliation{Black Hole Initiative at Harvard University, 20 Garden Street, Cambridge, MA 02138, USA}

\author[0000-0003-3664-963X]{Tiffany Nichols}
\affiliation{Department of History, Princeton University, Dickinson Hall, Princeton, NJ 08544, USA}

\author[0000-0002-5278-9221]{Dominic W. Pesce}
\affiliation{Center for Astrophysics $|$ Harvard \& Smithsonian, 60 Garden Street, Cambridge, MA 02138, USA}
\affiliation{Black Hole Initiative at Harvard University, 20 Garden Street, Cambridge, MA 02138, USA}

\author[0000-0001-9283-1191]{Ziri Younsi}
\affiliation{Mullard Space Science Laboratory, University College London, Holmbury St. Mary, Dorking, Surrey, RH5 6NT, UK}

%%%%%%%%%%%%%%%%%%%%%%%%%%%%%%%%%%%%
% Other direct or major contributors
% Also remaining ngEHT PO and Engineering Team
%%%%%%%%%%%%%%%%%%%%%%%%%%%%%%%%%%%%

\author[0000-0003-2966-6220]{Andrew Chael}
\affiliation{Princeton Gravity Initiative, Princeton University, Jadwin Hall, Princeton, NJ 08544, USA}

\author[0000-0002-2825-3590]{Koushik Chatterjee}
\affiliation{Black Hole Initiative at Harvard University, 20 Garden Street, Cambridge, MA 02138, USA}
\affiliation{Center for Astrophysics $|$ Harvard \& Smithsonian, 60 Garden Street, Cambridge, MA 02138, USA}

\author{Ryan Chaves}
\affiliation{Center for Astrophysics $|$ Harvard \& Smithsonian, 60 Garden Street, Cambridge, MA 02138, USA}

\author[0000-0001-7864-2458]{Juliusz Doboszewski}
\affiliation{Black Hole Initiative at Harvard University, 20 Garden Street, Cambridge, MA 02138, USA}
\affiliation{Lichtenberg Group for History and Philosophy of Physics, University of Bonn}

\author[0000-0003-0392-3604]{Richard Dodson}
\affiliation{ICRAR, M468, The University of Western Australia, 35 Stirling Hwy, Crawley WA 6009, Australia}

\author[0000-0002-9031-0904]{Sheperd~S.~Doeleman}
\affiliation{Center for Astrophysics $|$ Harvard \& Smithsonian, 60 Garden Street, Cambridge, MA 02138, USA}
\affiliation{Black Hole Initiative at Harvard University, 20 Garden Street, Cambridge, MA 02138, USA}

\author[0000-0002-1482-5682]{Jamee Elder}
\affiliation{Black Hole Initiative at Harvard University, 20 Garden Street, Cambridge, MA 02138, USA}
\affiliation{Lichtenberg Group for History and Philosophy of Physics, University of Bonn}

\author{Garret Fitzpatrick}
\affiliation{Center for Astrophysics $|$ Harvard \& Smithsonian, 60 Garden Street, Cambridge, MA 02138, USA}

\author{Kari Haworth}
\affiliation{Center for Astrophysics $|$ Harvard \& Smithsonian, 60 Garden Street, Cambridge, MA 02138, USA}

\author{Janice Houston}
\affiliation{Center for Astrophysics $|$ Harvard \& Smithsonian, 60 Garden Street, Cambridge, MA 02138, USA}

\author[0000-0002-5297-921X]{Sara Issaoun}
\altaffiliation{NASA Hubble Fellowship Program, Einstein Fellow}
\affiliation{Center for Astrophysics $|$ Harvard \& Smithsonian, 60 Garden Street, Cambridge, MA 02138, USA}

\author[0000-0001-9303-3263]{Yuri Y.~Kovalev}
\affiliation{Max-Planck-Institut f\"ur Radioastronomie, Auf dem H\"ugel 69, D-53121 Bonn, Germany}
\affiliation{Black Hole Initiative at Harvard University, 20 Garden Street, Cambridge, MA 02138, USA}

\author[0000-0001-7307-632X]{Aviad Levis}
\affiliation{California Institute of Technology, 1200 East California Boulevard, Pasadena, CA 91125, USA}

\author[0000-0001-7361-2460]{Rocco Lico}
\affiliation{Instituto de Astrof\'{\i}sica de Andaluc\'{\i}a-CSIC, Glorieta de la Astronom\'{\i}a s/n, E-18008 Granada, Spain}
\affiliation{INAF-Istituto di Radioastronomia, Via P. Gobetti 101, I-40129 Bologna, Italy}

\author[0000-0002-5780-0805]{Alexandru {Marcoci}}
\affiliation{Centre for the Study of Existential Risk, University of Cambridge, 16 Mill Lane, Cambridge, CB2 1SB, UK}

\author[0000-0002-2839-1387]{Niels~C.M.~{Martens}}
\affiliation{Freudenthal Institute, Utrecht University, the Netherlands}
\affiliation{Descartes Centre for the History and Philosophy of the Sciences and the Humanities, Utrecht University, the Netherlands}
\affiliation{Lichtenberg Group for History and Philosophy of Physics, University of Bonn}

\author[0000-0001-6920-662X]{Neil~M.~Nagar}
\affiliation{Astronomy Department, Universidad de Concepci\'on, Casilla 160-C, Concepci\'on, Chile}

\author{Aaron Oppenheimer}
\affiliation{Center for Astrophysics $|$ Harvard \& Smithsonian, 60 Garden Street, Cambridge, MA 02138, USA}

\author[0000-0002-7179-3816]{Daniel~C.~M.~Palumbo}
\affiliation{Center for Astrophysics $|$ Harvard \& Smithsonian, 60 Garden Street, Cambridge, MA 02138, USA}
\affiliation{Black Hole Initiative at Harvard University, 20 Garden Street, Cambridge, MA 02138, USA}

\author[0000-0001-5287-0452]{Angelo Ricarte}
\affiliation{Black Hole Initiative at Harvard University, 20 Garden Street, Cambridge, MA 02138, USA}
\affiliation{Center for Astrophysics $|$ Harvard \& Smithsonian, 60 Garden Street, Cambridge, MA 02138, USA}

\author[0000-0003-4871-9535]{Mar\'{\i}a\,J. Rioja}
\affiliation{ICRAR, M468, The University of Western Australia, 35 Stirling Hwy, Crawley WA 6009, Australia}
\affiliation{CSIRO Astronomy and Space Science, PO Box 1130, Bentley WA 6102, Australia}
\affiliation{Observatorio Astron\'omico Nacional (IGN), Alfonso XII, 3 y 5, 28014 Madrid, Spain}

\author[0000-0001-5461-3687]{Freek Roelofs}
\affiliation{Center for Astrophysics $|$ Harvard \& Smithsonian, 60 Garden Street, Cambridge, MA 02138, USA}
\affiliation{Black Hole Initiative at Harvard University, 20 Garden Street, Cambridge, MA 02138, USA}
\affiliation{Department of Astrophysics, Institute for Mathematics, Astrophysics and Particle Physics (IMAPP), Radboud University, P.O. Box 9010, 6500 GL Nijmegen, The Netherlands}

\author[0000-0002-1976-6407]{Ann C. {Thresher}}
\affiliation{McCoy Family Center for Ethics, Stanford University}

\author[0000-0003-3826-5648]{Paul Tiede}
\affiliation{Center for Astrophysics $|$ Harvard \& Smithsonian, 60 Garden Street, Cambridge, MA 02138, USA}
\affiliation{Black Hole Initiative at Harvard University, 20 Garden Street, Cambridge, MA 02138, USA}

\author[0000-0002-4603-5204]{Jonathan Weintroub}
\affiliation{Center for Astrophysics $|$ Harvard \& Smithsonian, 60 Garden Street, Cambridge, MA 02138, USA}
\affiliation{Black Hole Initiative at Harvard University, 20 Garden Street, Cambridge, MA 02138, USA}

\author[0000-0002-8635-4242]{Maciek Wielgus}
\affiliation{Max-Planck-Institut f\"ur Radioastronomie, Auf dem H\"ugel 69, D-53121 Bonn, Germany}

% \author{The ngEHT Project}

\begin{abstract}
The Event Horizon Telescope (EHT) has led to the first images of a supermassive black hole, revealing the central compact objects in the elliptical galaxy M87 and the Milky Way. Proposed upgrades to this array through the next-generation EHT (ngEHT) program would sharply improve the angular resolution, dynamic range, and temporal coverage of the existing EHT observations. These improvements will uniquely enable a wealth of transformative new discoveries related to black hole science, extending from event-horizon-scale studies of strong gravity to studies of explosive transients to the cosmological growth and influence of supermassive black holes. Here, we present the key science goals for the ngEHT and their associated instrument requirements, both of which have been formulated through a multi-year international effort involving hundreds of scientists worldwide.\\ 
\end{abstract}
 
\input{body}

\acknowledgments{
We are pleased to acknowledge the hundreds of scientists and engineers worldwide who have contributed to the ngEHT science case and have helped define the associated instrument requirements. In particular, we are grateful to the panelists of the ngEHT Science Requirements Review: Jim Moran, Mariafelicia De Laurentis, Eric Murphy, Geoff Bower, Katherine Blundell, and Randy Iliff. The detailed feedback from this review substantially sharpened the science goals and informed the specification of threshold and objective requirements. We also thank the EHTC internal referee, Laurent Loinard, the MPIfR internal referee, Eduardo Ros, and the anonymous referees for their detailed feedback and suggestions, which significantly improved the manuscript. 
The ngEHT design studies are funded by National Science Foundation grants AST-1935980 and AST-2034306 and the Gordon and Betty Moore Foundation (GBMF-10423). This work was supported by the Black Hole Initiative at Harvard University, which is funded by grants from the John Templeton Foundation and the Gordon and Betty Moore Foundation to Harvard University. 
This work was supported by Volkswagen Foundation, VILLUM Foundation (grant no.\ VIL37766) and the DNRF Chair program (grant no.\ DNRF162) by the Danish National Research Foundation. We acknowledge financial support provided under the European Union's H2020 ERC Advanced Grant ``Black holes: gravitational engines of discovery'' grant agreement no.\ Gravitas–101052587. NCMM acknowledges support from the European Union's Horizon Europe research and innovation programme for the funding received under the Marie Sk\l{}odowska-Curie grant agreement No.~101065772 (PhilDarkEnergy) and the ERC Starting Grant agreement No.~101076402 (COSMO-MASTER). Views and opinions expressed are however those of the authors only and do not necessarily reflect those of the European Union or the European Research Council. Neither the European Union nor the granting authority can be held responsible for them. NN acknowledges funding from TITANs NCN19-058 and Fondecyt 1221421.
}

\bibliography{main.bbl}

\end{document}

%% file: body.tex
%%%%%%%%%%%%%%%%%%%%%%%%%
\section{Introduction}
%%%%%%%%%%%%%%%%%%%%%%%%%
The Event Horizon Telescope (EHT) has produced the first images of the supermassive black holes (SMBHs) in the M87 galaxy \citep[][hereafter \m87~I-VIII]{EHTC_M87_I,EHTC_M87_II,EHTC_M87_III,EHTC_M87_IV,EHTC_M87_V,EHTC_M87_VI,EHTC_M87_VII,EHTC_M87_VIII} and at the center of the Milky Way \citep[][hereafter \sgra~I-VI]{EHTC_SgrA_I,EHTC_SgrA_II,EHTC_SgrA_III,EHTC_SgrA_IV,EHTC_SgrA_V,EHTC_SgrA_VI}. Interpretation of the EHT results for \sgra and \m87 has relied heavily upon coordinated multi-wavelength campaigns spanning radio to gamma-rays \citepalias[\citealt{EHT_MWL};][]{EHTC_SgrA_II}. 
In addition, the EHT has made the highest resolution images to date of the inner jets of several nearby Active Galactic Nuclei (AGN), demonstrating the promise of millimeter VLBI in making major contributions to the studies of relativistic radio jets launched from SMBHs (\citealt{Kim_2020,Janssen_2021,Issaoun_2022,Jorstad_2023}).

The EHT results were achieved using the technique of very-long-baseline interferometry (VLBI). In this approach, radio signals are digitized and recorded at a collection of telescopes; the correlation function between every pair of telescopes is later computed offline, with each correlation coefficient sampling one Fourier component of the sky image with angular frequency given by the dimensionless vector baseline (measured in wavelengths) projected orthogonally to the line of sight \citep{TMS}. 
The EHT observations at 230\,GHz are the culmination of pushing VLBI to successively higher frequencies across decades of development \citep[e.g.,][]{Padin_1990,Krichbaum_1998,Doeleman_2008}, giving a diffraction-limited angular resolution of ${\sim}20\,\mu{\rm as}$ \citep[for a review of mm-VLBI, see][]{Boccardi_2017}. For comparison, the angular diameter of the lensed event horizon -- the BH ``shadow'' -- is $\theta_{\rm sh} \approx 10 G M/(c^2 D)$, where $G$ is the gravitational constant, $c$ is the speed of light, $M$ is the BH mass, and $D$ is the BH distance \citep{Bardeen_1973,Luminet_1979,Falcke_2000,deVries_2000}. For \m87, $\theta_{\rm sh} \approx 40\,\mu{\rm as}$; for \sgra, $\theta_{\rm sh} \approx 50\,\mu{\rm as}$. 

\begin{figure*}[t]
\begin{center}
\includegraphics[width=1.00\textwidth]{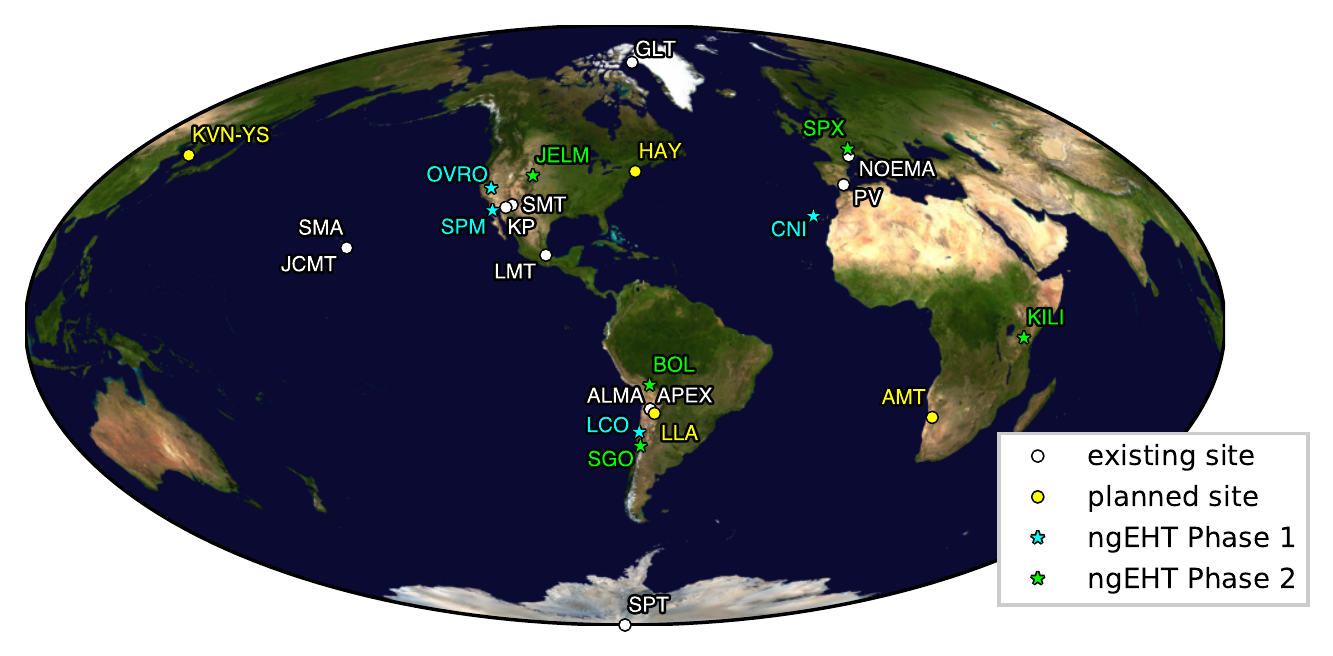}
\caption{
Distribution of EHT and ngEHT sites around the globe.  Sites that have joined EHT campaigns are shown in white \citepalias[see][]{EHTC_M87_II}, additional ngEHT Phase-1 sites are shown in cyan, and ngEHT Phase-2 sites are shown in green. 
Three of the EHT sites have joined since its initial observing campaign in 2017: the 12-m Greenland Telescope \citep[GLT;][]{Inoue_2014}, the 12-m Kitt Peak Telescope (KP), the Northern Extended Millimeter Array (NOEMA) composed of twelve 15-m dishes. 
Several other existing or upcoming sites that plan to join EHT/ngEHT observations are shown in yellow: the 37-m Haystack Telescope \citep[HAY;][]{Kauffmann_2023}, the 21-m Yonsei Radio Observatory of the Korea VLBI Network \citep[KVN-YS;][]{Asada_2017}, the 15-m Africa Millimetre Telescope \citep[AMT;][]{Backes_2016}, and the 12-m Large Latin American Millimeter Array \citep[LLA;][]{Romero_2020}. For additional details on the planned ngEHT specifications, see \citet{ngEHT_refarray}. 
}
\label{fig:globe}
\end{center}
\end{figure*}

\begin{figure*}[t]
\begin{center}
\includegraphics[width=0.9\textwidth]{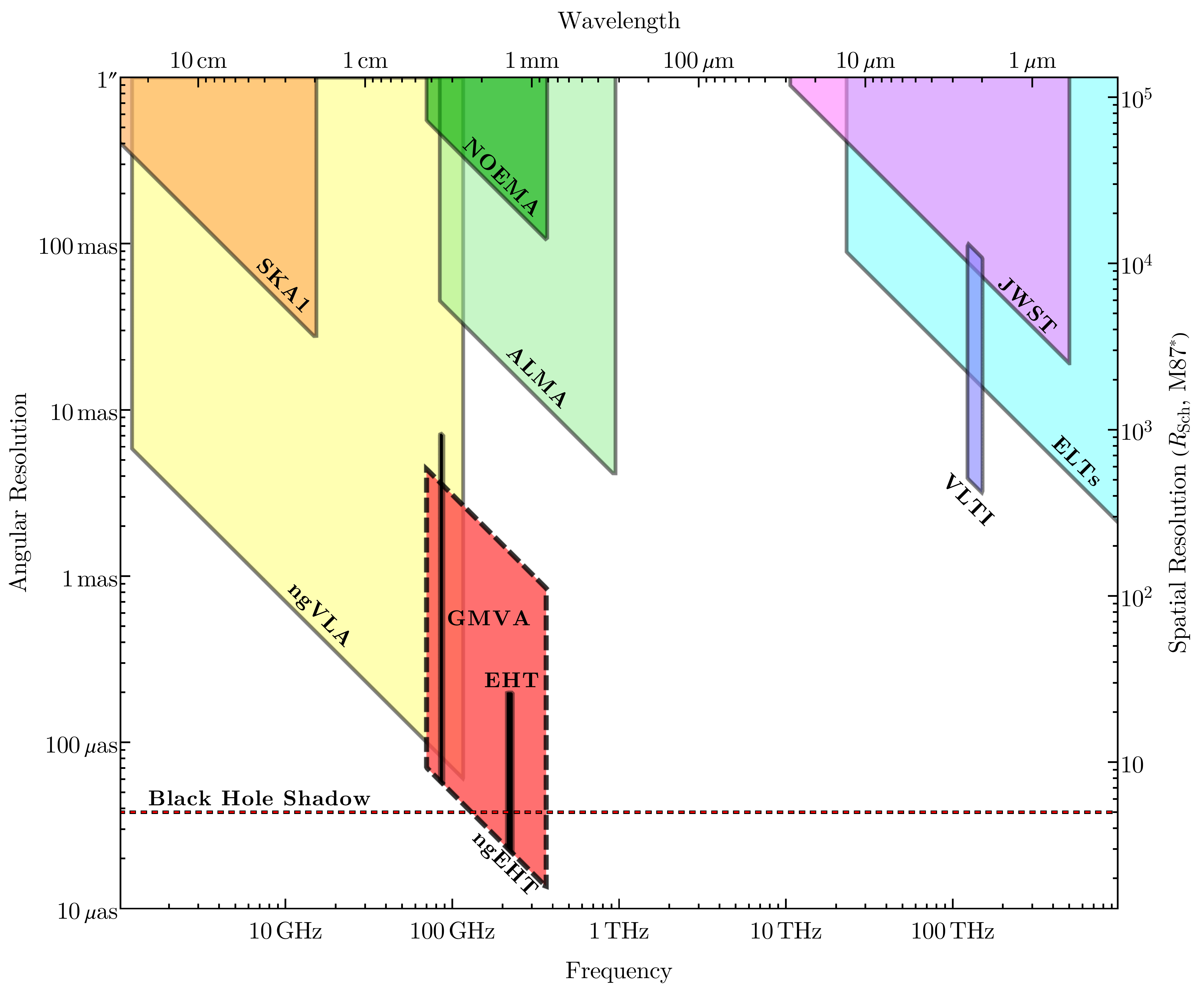}
\caption{
Range of observing frequency and angular resolution for selected current and upcoming facilities, from radio to the infrared. The ngEHT can achieve an imaging angular resolution that is significantly finer than any other planned facility or experiment. The ngEHT also envisions simultaneous multi-band observations, extending from 86 to 345\,GHz, which will significantly expand the frequency coverage of currently published EHT data (black filled region). 
Figure adapted from \citet{Selina_2018}.
}
\label{fig:ngeht_facility_comparison}
\end{center}
\end{figure*}

Despite the remarkable discoveries of the EHT, they represent only the first glimpse of the promise of horizon-scale imaging studies of BHs and of high-frequency VLBI more broadly. In particular, the accessible science in published EHT results is severely restricted in several respects:
\begin{itemize}
    \item {\it EHT images are effectively monochromatic.} The currently published EHT measurements sample only 4\,GHz of bandwidth, centered on 228\,GHz. BH images are expected to have a complex structure in frequency, with changing synchrotron emissivity, optical depth, and Faraday effects, making multi-frequency studies a powerful source of physical insight \citep[see, e.g.,][]{Moscibrodzka_2017,Ricarte_2020,Chael_2022,Ricarte_2023_pol}. The EHT has successfully completed commissioning observations at 345\,GHz \citep{Crew_2023}, which is now a standard observing mode. However, 345\,GHz observations will be strongly affected by atmospheric absorption, severely affecting sensitivity and likely restricting detections to intermediate baseline lengths among the most sensitive sites \citep[e.g.,][]{Roelofs_2023}. 
    \item {\it EHT images have severely limited image dynamic range.} Current EHT images are limited to a dynamic range of only ${\sim}10$ \citepalias{EHTC_M87_IV,EHTC_SgrA_III}, providing only modest information about image signatures that are related to the horizon and 
    limiting the ability to connect the event-horizon-scale images to their relativistic jets seen until now only at larger scales, via lower wavelength observations.\footnote{Since its first observing campaign, three sites have joined the EHT (see \autoref{fig:globe}). These additions are expected to substantially improve upon the dynamic range of published EHT images.} 
    For comparison, VLBI arrays operating at centimeter wavelengths routinely achieve a dynamic range of ${\sim}10^4$ on targets such as \m87 \citep[e.g.,][]{Walker_2018}. 
    \item {\it EHT observations have only marginally resolved the rings in \sgra and \m87.} The EHT only samples a few resolution elements across the images. For instance, the EHT has only determined an upper limit on the thickness of the \m87 ring \citepalias{EHTC_M87_VI}, and the azimuthal structure of the rings in both sources is poorly constrained. 
    \item {\it EHT images cannot yet study the dynamics of \m87 or \sgra.}
    The gravitational timescale is $t_{\rm g} \equiv GM/c^3 \approx 9\,{\rm hours}$ for \m87 and is $t_{\rm g} \approx 20\,{\rm s}$ for \sgra. In each source, the expected evolution timescale is ${\sim}50\,t_{\rm g}$ \citep[e.g.,][]{Wielgus_2020} ---  approximately $20\,{\rm days}$ for \m87 and $20\,{\rm minutes}$ for \sgra. 
    Current EHT campaigns consist of sequential observing nights extending for only ${\sim}$1~week, which is too short to study the dynamical evolution of \m87. Moreover, the current EHT baseline coverage is inadequate to meaningfully constrain the rapid dynamical evolution of \sgra, which renders standard Earth-rotation synthesis imaging inapplicable \citepalias{EHTC_SgrA_III,EHTC_SgrA_IV}.
\end{itemize}
In short, published EHT images of \m87 and \sgra currently sample only $5 \times 5$ spatial resolution elements, a single spectral resolution element, and a single temporal resolution element (snapshot for \m87; time-averaged for \sgra).

\begin{figure*}[t]
\begin{center}
\includegraphics[width=0.9\textwidth]{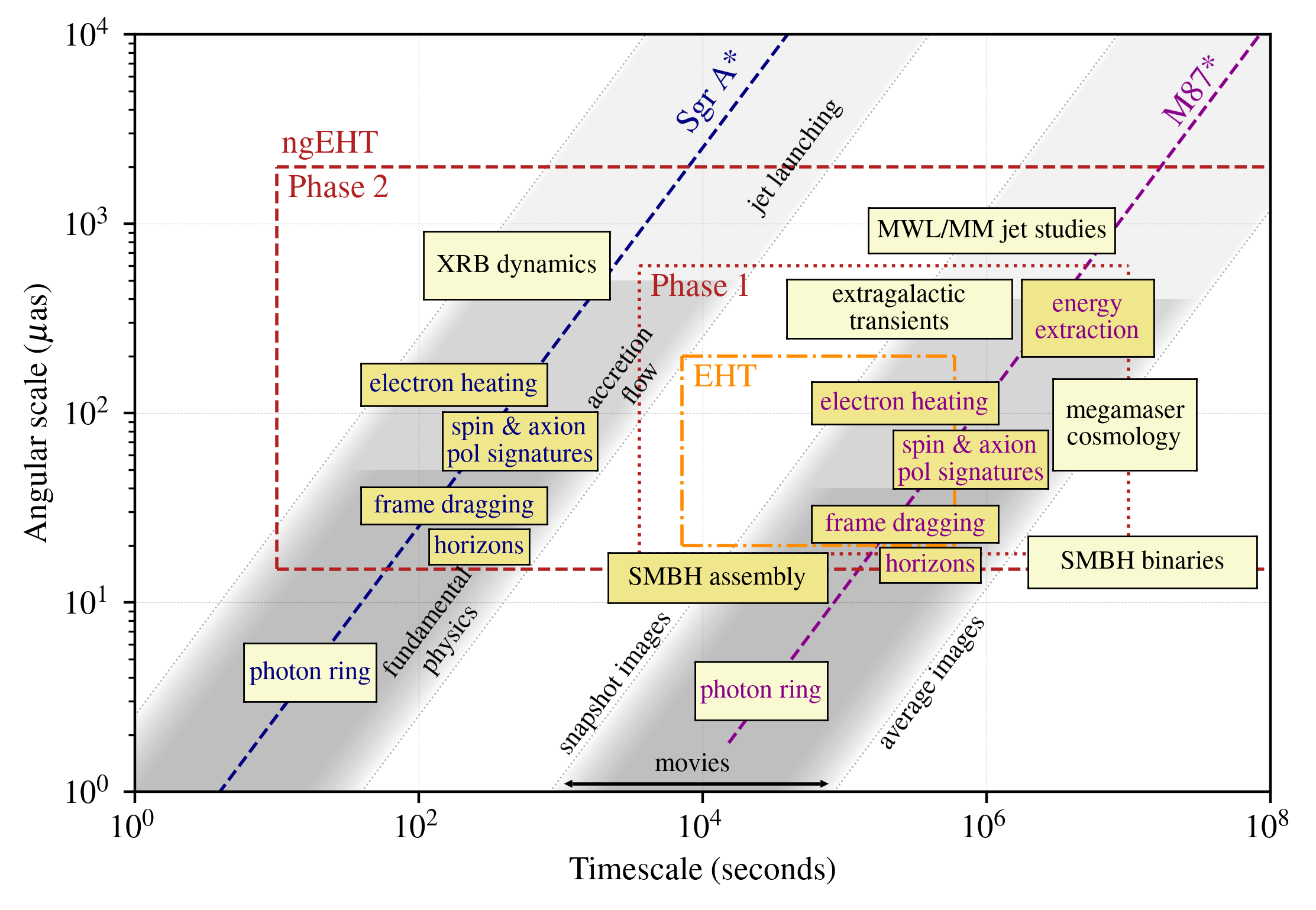}
\caption{
Comparison of image angular resolutions and timescales accessible to the EHT and ngEHT and the associated scientific opportunities. For \m87 and \sgra, the ranges of angular resolution and timescale needed to study the three primary domains -- fundamental physics, accretion, and jet launching -- are indicated with the tilted shaded regions. These shaded regions are centered on the resolution-timescale for each source determined by the speed of light ($c t = D \theta$). Snapshot images require an array to form images on these timescales or shorter; average images require an array to form images over significantly longer timescales; movies require that an array can form images of the full range of timescales from snapshots to averages. 
The primary difference in \m87 and \sgra is the factor of ${\sim}1500$ difference in the SMBH mass, which sets the system timescale. In contrast, the relevant angular scales in these systems are determined by the mass-to-distance ratio, which only differs by ${\sim}20\%$ for these two SMBHs. 
The approximate resolution-timescale pair to study each of the ngEHT Key Science Goals is indicated with the inset labeled boxes. Goals associated with \sgra or \m87 are colored in blue or purple, respectively.
}
\label{fig:resolution_timescale}
\end{center}
\end{figure*}

The next-generation EHT (ngEHT) is a project to design and build a significantly enhanced EHT array through the upgrade, integration, or deployment of additional stations \citep[e.g.,][]{Backes_2016,Asada_2017,Romero_2020,Raymond_2021,Bustamante_2023,Kauffmann_2023,Yu_2023,Akiyama_2023}, the use of simultaneous observations at three observing frequencies \citep{Issaoun_2023,Rioja_2023,Jiang_2023}, and observations that extend over several months or more with a dense coverage in time \citep{Doeleman_2019}. The ngEHT currently envisions two primary development phases. In Phase~1, the ngEHT will add 3 or more dedicated telescopes to the current EHT, with primarily dual-band (230/345\,GHz) observations over ${\sim}3$ months per year.\footnote{In contrast, most telescopes of the present EHT are astronomical facilities that only commit a small fraction of their total observing time to VLBI.} 
In Phase~2, the ngEHT will add 5 or more additional dedicated telescopes, with simultaneous tri-band capabilities (86/230/345\,GHz) at most sites and observations available year-round. The new ngEHT antennas are expected to have relatively modest diameters (6-10\,m), relying on wide recorded bandwidths, strong baselines to large existing apertures, and long integrations enabled through simultaneous multi-band observations to achieve the needed baseline sensitivity. \autoref{fig:globe} shows candidate ngEHT sites in each phase.

These developments will sharply improve upon the performance of the EHT. Relative to other premier and planned facilities that target high-resolution imaging (such as the SKA, ngVLA, ALMA, and ELTs), the defining advantage of the EHT is its unmatched angular resolution. However, relative to the imaging capabilities of the current EHT, the defining improvements of the ngEHT images will be in accessing \emph{larger} angular scales through the addition of \emph{shorter} interferometric baselines than those of the present array, and in expanding the simultaneous frequency coverage. In addition, the ngEHT will extend accessible timescales of the current EHT by ${\sim}5$ orders of magnitude, enabling dynamic analysis with the creation of movies of \sgra (through improved ``snapshot'' imaging on ${\sim}$minute timescales) and AGN including \m87 (through densely sampled monitoring campaigns that extend from months to years). \autoref{fig:ngeht_facility_comparison} and \autoref{fig:resolution_timescale} summarize these improvements.  

To guide its design, the ngEHT has developed a set of Key Science Goals over the past two years, with contributions from hundreds of scientists. 
This process has included three international meetings,\footnote{\url{https://www.ngeht.org/ngeht-meeting-2021}}\textsuperscript{,}\footnote{\url{https://www.ngeht.org/ngeht-meeting-november-2021}}\textsuperscript{,}\footnote{\url{https://www.ngeht.org/ngeht-meeting-june-2022}}, a Science Requirements Review (focused on identifying the most significant ngEHT science drivers), and a System Requirements Review (focused on identifying the associated instrument requirements). 
In addition, the ngEHT project has convened focused workshops on major topics, including assessing the motivation for adding 86\,GHz capabilities to the ngEHT design to leverage phase transfer techniques,\footnote{\url{https://www.ngeht.org/broadening-horizons-2022}}, environmental and cultural issues related to ethical telescope siting, and the role of History, Philosophy, and Culture in the ngEHT \citep[see \autoref{sec:HPC} and][hereafter \citetalias{HPC}]{HPC}. A series of papers present many aspects of these science cases in greater depth in a special issue of {\it Galaxies}.\footnote{\url{https://www.mdpi.com/journal/galaxies/special_issues/ngEHT_blackholes}} 

In this paper, we summarize the Key Science Goals of the ngEHT and associated instrument requirements. We begin by discussing specific scientific objectives, organized by theme, in \autoref{sec:Science_Drivers}. We then summarize the prioritization and aggregated requirements of these science cases together with a condensed version of the ngEHT Science Traceability Matrix (STM) in \autoref{sec:Summary}. Details on the ngEHT concept, design, and architecture are presented in a companion paper \citep{ngEHT_refarray}.

\section{Key Science Goals of the ngEHT}
\label{sec:Science_Drivers}

The ngEHT Key Science Goals were developed across eight Science Working Groups (SWGs). These goals span a broad range of targets, spatial scales, and angular resolutions (see \autoref{fig:resolution_timescale}). 
We now summarize the primary recommendations from each of these working groups: Fundamental Physics (\autoref{sec:Fundamental_Physics}), Black Holes \& their Cosmic Context (\autoref{sec:BHCC}), Accretion (\autoref{sec:Accretion}), Jet Launching (\autoref{sec:Jet_Launching}), Transients (\autoref{sec:Transients}), New Horizons (\autoref{sec:New_Horizons}), Algorithms \& Inference (\autoref{sec:AI}), and History, Philosophy, \& Culture (\autoref{sec:HPC}).

\begin{figure*}[t]
    \centering
    \includegraphics[width=\textwidth]{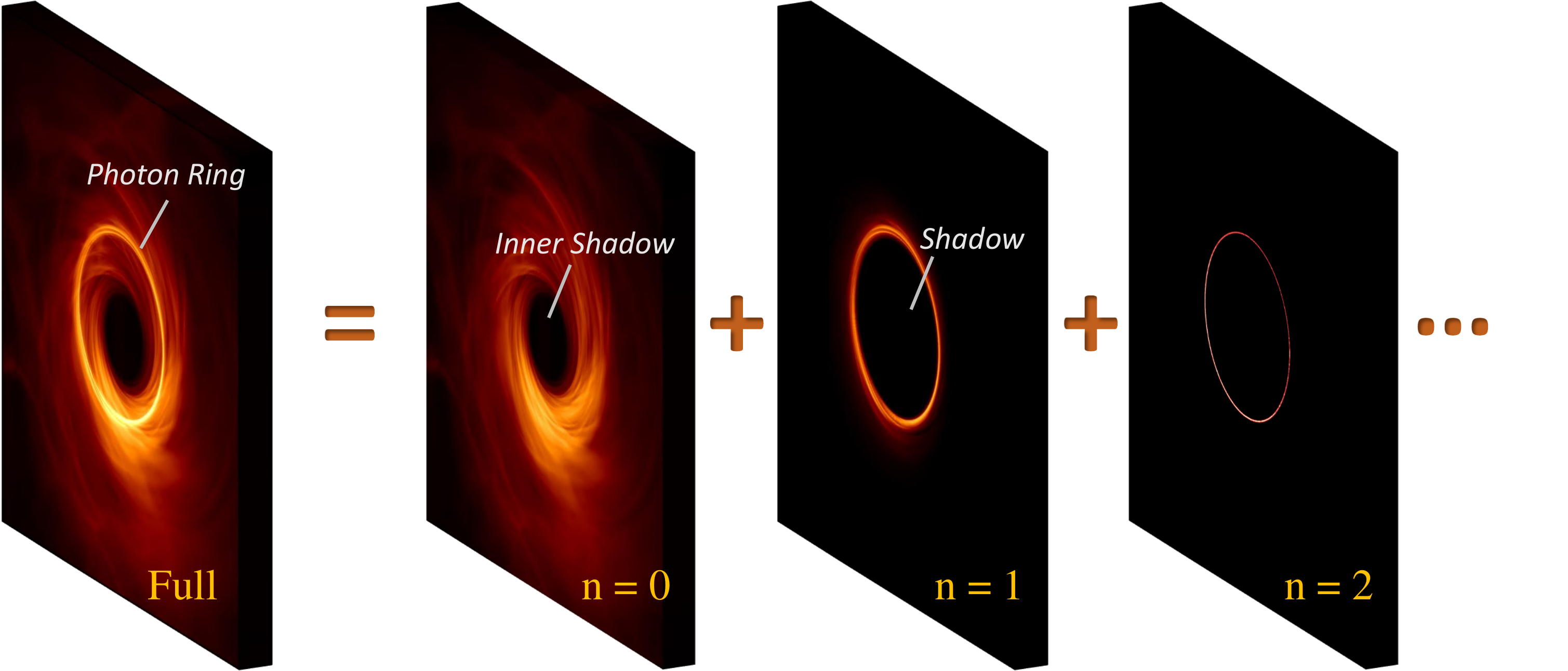}
    \caption{BH images display a series of distinctive relativistic features such as the BH apparent ``shadow'' \citep[e.g.,][]{Falcke_2000}, ``inner shadow'' \citep[e.g.,][]{Chael_2021}, and ``photon ring'' \citep[e.g.,][]{Johnson_2020}.
    }
\label{fig:PhotonRingStack}
\end{figure*}

%%%%%%%%%%%%%%%%%%%%%%%%%%%%%%%%%%%
\subsection{Fundamental Physics}
\label{sec:Fundamental_Physics}
%%%%%%%%%%%%%%%%%%%%%%%%%%%%%%%%%%
BHs are an extraordinary prediction of general relativity: the most generic and simple macroscopic objects in the Universe. Among astronomical targets, BHs are unique in their ability to convert energy into electromagnetic and gravitational radiation with remarkable efficiency \citep[e.g.,][]{Penrose_1969,Shakura_Sunyaev_1973,Blandford_Znajek_1977,Yuan_Narayan_2014}. Meanwhile, the study of BH stability and dynamics challenges our knowledge of partial differential equations, of numerical methods, and of the interplay between quantum field theory and the geometry of spacetime. The BH information paradox \citep[e.g.,][]{Harlow_2016} and the existence of unresolved singularities in classical general relativity \citep[e.g.,][]{Penrose_1969,Senovilla_2015} point to deep inconsistencies in our current understanding of gravity and quantum mechanics. It is becoming clear that the main conceptual problems in BH physics hold the key to many current open foundational issues in theoretical physics. 

Astrophysical BH systems are therefore an extraordinary test-bed for fundamental physics, although their extreme compactness renders them observationally elusive. Matter moving in the vicinity of an event horizon is subject to both extreme (thermo)dynamical conditions and intense gravitational fields, thereby providing a unique laboratory for the study of physical processes and phenomena mediated by the strongest gravitational fields in the Universe. Furthermore, by understanding the properties of matter and polarised electromagnetic radiation in this highly-nonlinear (and dynamical) regime, we can probe the underlying spacetime geometry of BHs and perform new tests of general relativity. Key to studying physics near the horizon is the capability to resolve, accurately extract, and precisely measure different features in BH images (see \autoref{fig:PhotonRingStack}). These image features can be periodic (e.g., oscillating fields), transient (e.g., reconnective processes and flares), persistent (the photon ring), or stochastic about a mean (e.g., polarization spiral patterns).

Previous measurements of \m87 and \sgra from the EHT provide compelling evidence for supermassive compact objects. The ngEHT has the capability to elevate existing EHT probes of the strong-field regime. We now describe four key science goals that target foundational topics in fundamental physics: studies of horizons (\autoref{sec:Horizon_Physics}), measurements of SMBH spin (\autoref{sec:Spin}), studies of a BH photon ring (\autoref{sec:Photon_Ring}), and constraints on ultralight boson fields (\autoref{sec:axions}). For a comprehensive discussion of these topics, see \citet{FP_Review}.

\begin{figure}[t]
    \centering
    \includegraphics[width=\columnwidth]{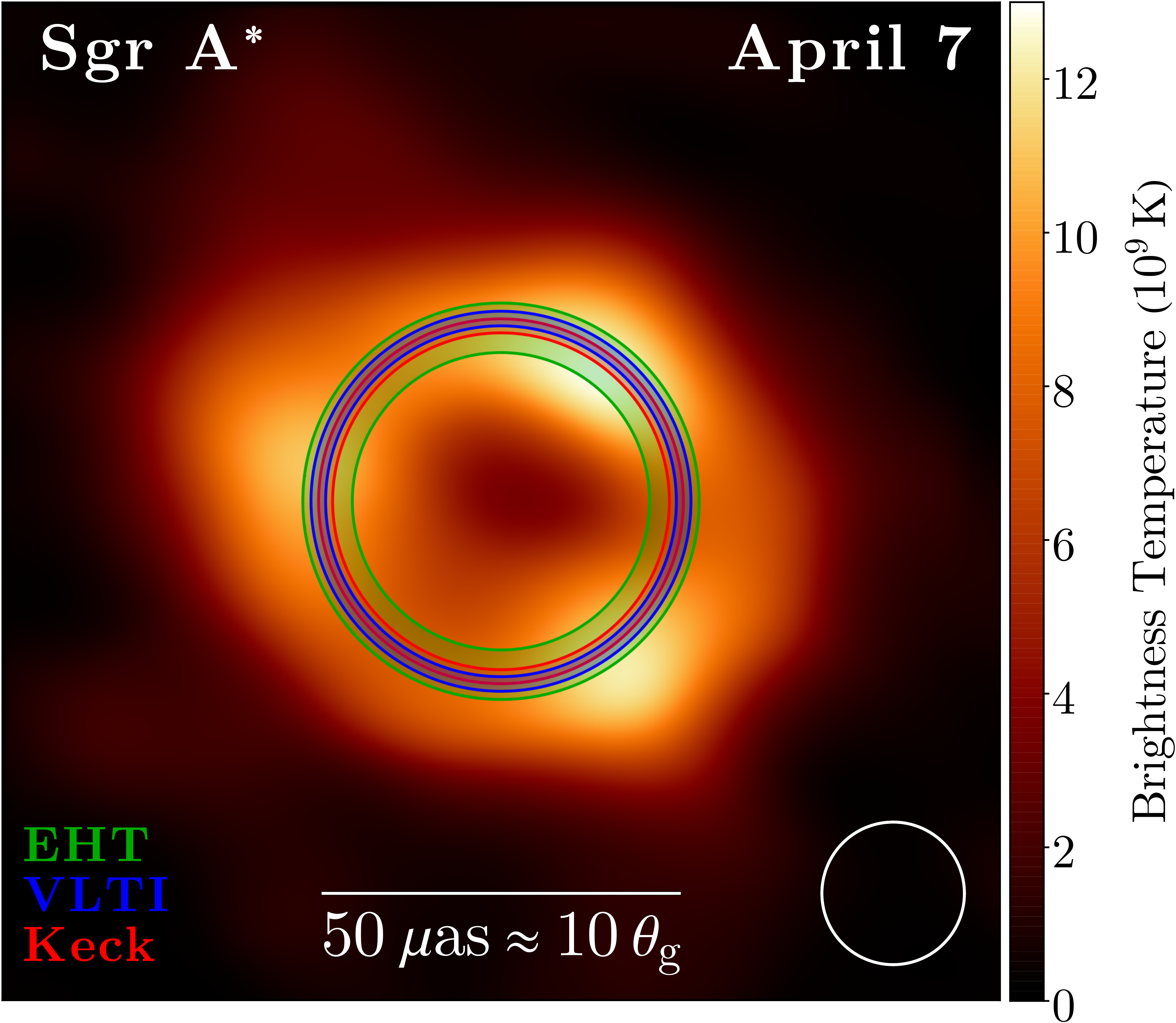}
    \caption{
    EHT representative average image of \sgra using data from April~7, 2017 \citepalias{EHTC_SgrA_III}. The white circle in the lower-right shows a $20\,\mu{\rm as}$ beam that gives the approximate EHT resolution. The overlaid annuli show the predicted ranges of the \sgra critical curve using measurements of resolved stellar orbits using the VLTI \citep[blue;][]{GRAVITY_2022} and Keck \citep[red;][]{Do_2019}; the ranges are dominated by the potential variation in size with spin, $d_{\rm sh} = (9.6 - 10.4) \theta_{\rm g}$ \citep{Bardeen_1973,Takahashi_2004}. The green annulus shows the estimated range ($\pm 1 \sigma$) of the critical curve using EHT measurements, which is consistent with these predictions \citepalias{EHTC_SgrA_VI}. However, because of the limited baseline coverage of the EHT, key image features such as the azimuthal brightness around the ring and the depth and shape of the central brightness depression are only weakly constrained with current observations. 
    }
\label{fig:EHT_SgrA}
\end{figure}

%%%%%%%%%%%%%%%%%%%%%%%%%%%%%%%%%%%%%%%%%%%%%%%%%%%%%%
\subsubsection{Existence and properties of horizons}
\label{sec:Horizon_Physics}
%%%%%%%%%%%%%%%%%%%%%%%%%%%%%%%%%%%%%%%%%%%%%%%%%%%%%%%
The formation of horizons (regions of spacetime that trap light) as gravitational collapse unfolds is one of the main and outstanding predictions of classical general relativity. Robust singularity theorems assert that BH interiors are also regions of breakdown of the classical Einstein equations, while quantum field theory is still associated with conundrums in the presence of horizons. Testing the existence and properties of horizons is therefore a key strong-field test of general relativity \citep{Carballo_2018,Cardoso_Pani_2019}. In astronomical terms, a horizon would be characterised by a complete absence of emission. It is clear that quantitative discussions of horizon physics will be strongly influenced both by the error in observational images and the modelling of matter and (spacetime) geometry at the core of simulated images.

For example, many models, especially those with spherical accretion onto BHs, tend to exhibit a pronounced apparent ``shadow'' \citep[e.g.,][]{Jaroszynski_Kurpiewski_1997,Falcke_2000,Narayan_2019,Younsi_2023}. This feature shows a sharp brightness gradient at the boundary of the ``critical curve'' that corresponds to the boundary of the observer's line of sight into the BH. In contrast, models in which the emission is confined to a thin disk that extends to the horizon show a sharp brightness gradient in a smaller feature, the ``inner shadow,'' which corresponds to the direct lensed image of the equatorial horizon \citep{Luminet_1979,Dokuchaev_2019,Chael_2021}. The inner shadow gives the observer's line of sight into the BH that is unobscured by the equatorial emitting region.

\begin{figure*}[t]
    \centering
    \includegraphics[width=\textwidth]{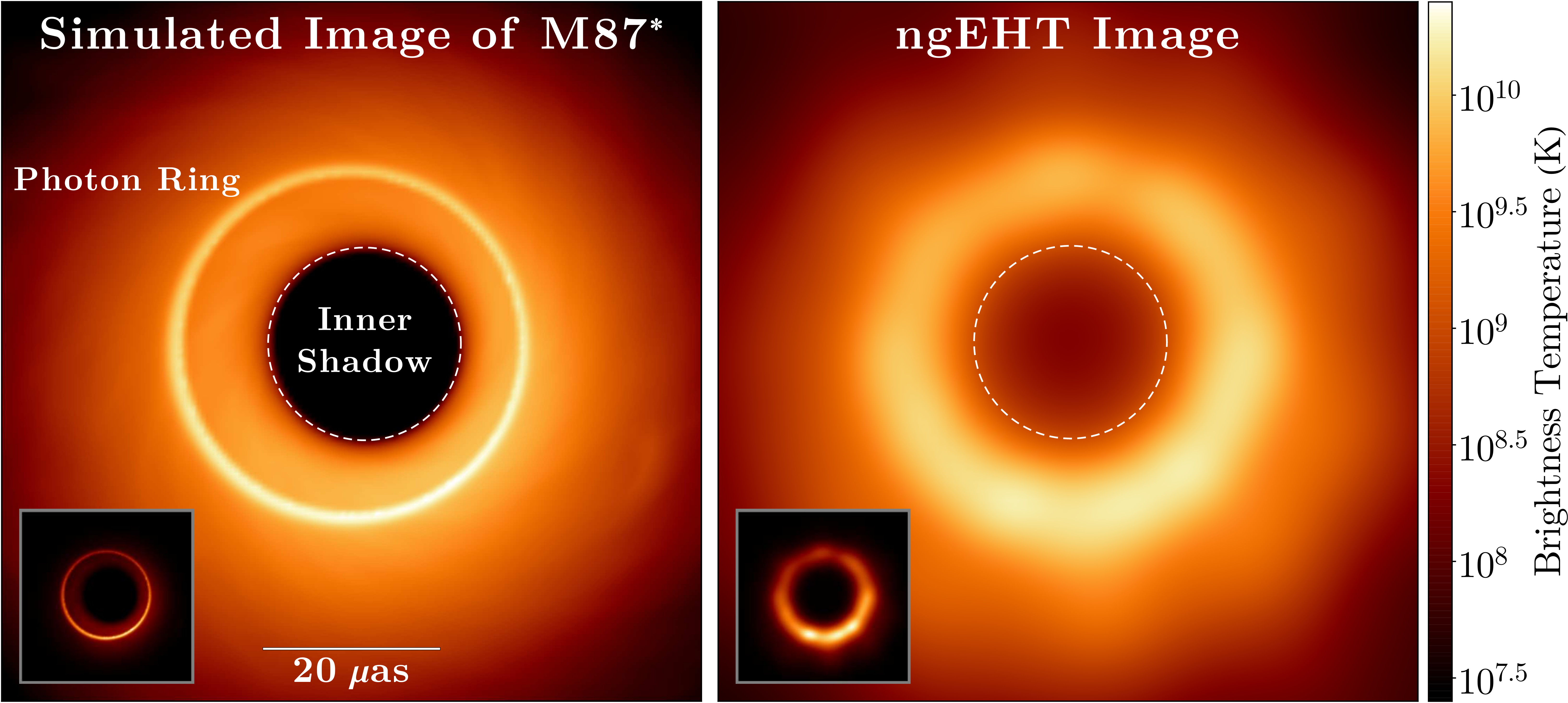}
    \caption{
    Accessing signatures of the event horizon with the ngEHT. Each panel shows an image on a logarithmic scale, with an inset shown with a linear scale. 
    The left panel shows a time-averaged simulated image of \m87, which shows a prominent photon ring and inner shadow. The right panel shows a reconstructed ngEHT image using the Bayesian VLBI analysis package \comrade \citep{Tiede_2022comrade} applied to simulated ngEHT phase-1 observations. The ngEHT provides both the angular resolution and dynamic range required to identify the deep brightness depression produced by the inner shadow in this simulated image.
    % Accessing signatures of the event horizon with the ngEHT. Each panel shows an image on a logarithmic scale, with an inset shown with a linear scale. 
    % The left panel shows a time-averaged simulated image of \m87, which shows a prominent photon ring and inner shadow. The center panel shows the current EHT image of \m87, which has severely limited dynamic range \citepalias{EHTC_M87_IV}. The right panel shows a reconstructed ngEHT image using the Bayesian VLBI analysis package \comrade \citep{Tiede_2022comrade} applied to simulated ngEHT phase-1 observations. The ngEHT provides both improved angular resolution and dynamic range, which identify the deep brightness depression produced by the inner shadow in this simulated image.
    }
\label{fig:Image_Reconstructions}
\end{figure*}

Hence, BHs can give rise to a rich array of distinctive image features, but studies of horizons through imaging must account for potential degeneracies between the properties of the spacetime and those of the emitting material. Firm conclusions from imaging with the ngEHT will require significant improvements in both the image dynamic range and angular resolution of current EHT images, which have primarily demonstrated consistency with predictions of the Kerr metric (see \autoref{fig:EHT_SgrA}) and order-unity constraints on potential violations of general relativity (see, e.g., \citealt{Psaltis_2020,Kocherlakota_2021}; \citetalias{EHTC_SgrA_VI}). To leading order, the image dynamic range of the ngEHT will determine the luminosity of the features that can be studied, while the angular resolution will determine the size of the features that can be studied. Hence, quantitative statements about the existence of horizons will be primarily influenced by the dynamic range that can be achieved, while quantitative properties of the spacetime will be determined by the angular resolution~\citep{Vincent:2020dij,Carballo-Rubio:2022aed}. \autoref{fig:Image_Reconstructions} shows an example of the improvement in both quantities that is possible using the ngEHT, enabling new studies of image signatures of the horizon. 
For additional discussion of potential ngEHT constraints on exotic horizonless spacetimes such as naked singularities and (non-hidden) wormholes, see \citet{FP_Review}. 
In addition to the necessity of image improvements, multi-frequency studies will be imperative to securely disentangle properties of the emission (which  are chromatic) from features associated with the lensed horizon (which is achromatic). For all studies of horizons through imaging with the ngEHT, \m87 and \sgra will be the primary targets because of their large angular sizes.

%%%%%%%%%%%%%%%%%%%%%%%%%%%%%%%%%%%%%%%%%%%%%
\subsubsection{Measuring the spin of a SMBH}
\label{sec:Spin}
%%%%%%%%%%%%%%%%%%%%%%%%%%%%%%%%%%%%%%%%%%%%%
Astrophysical BHs are expected to be completely characterized by their mass and angular momentum \citep{Robinson_1975,Gibbons_1975,Blandford_Znajek_1977}. Estimates of a SMBH spin through direct imaging would provide an invaluable complement to other techniques, such as the X-ray reflection method \citep[see, e.g.,][]{Reynolds_2021}. However, the current EHT measurements provide only marginal, model-dependent constraints on the spins of \m87 and \sgra \citepalias{EHTC_M87_V,EHTC_M87_VIII,EHTC_SgrA_V}. 

\begin{figure*}
    \centering
    \includegraphics[width=\textwidth]{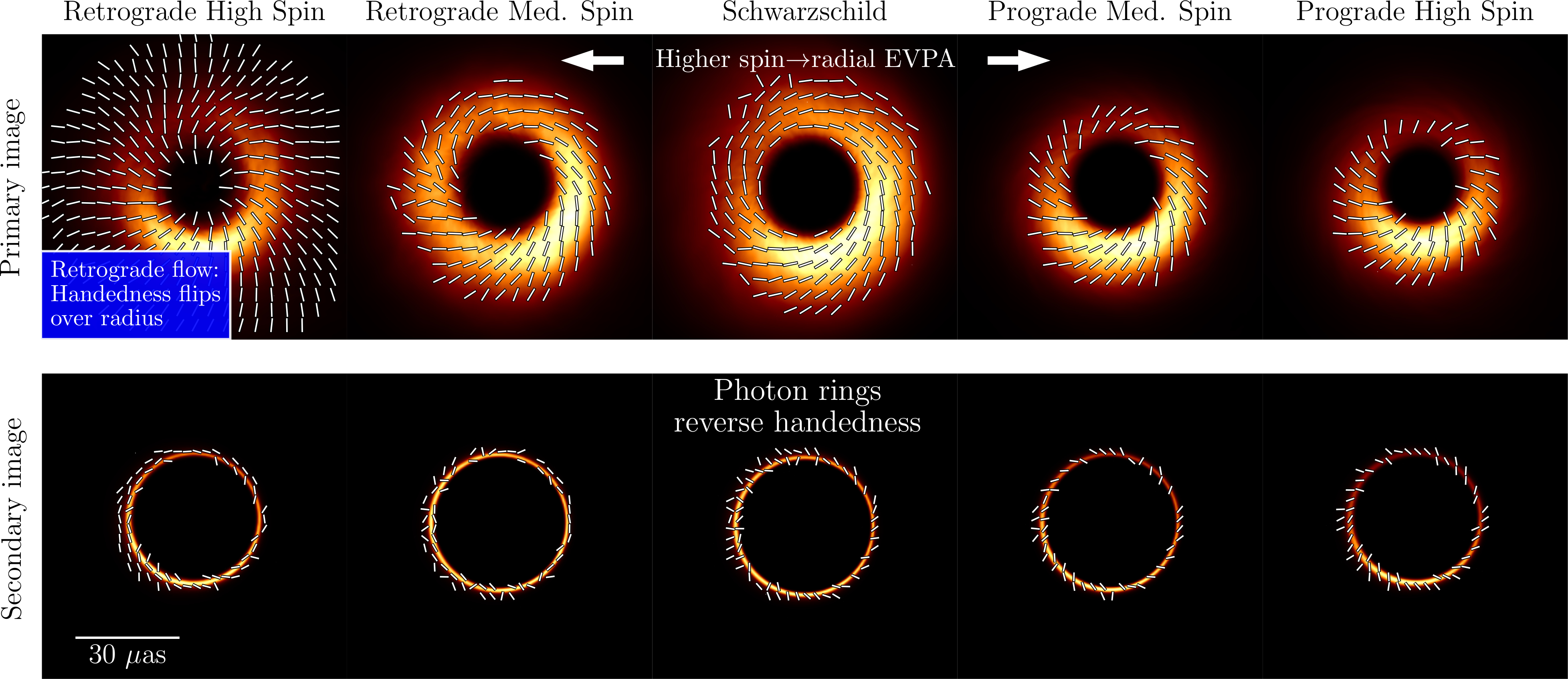}
    \caption{Summary of spin signatures in polarized images of time-averaged GRMHD simulations. In each panel, color indicates brightness and ticks show linear polarization direction. Rows show time-averaged primary (top) and secondary (bottom) images from MAD GRMHD simulations of \m87; columns show varying BH spin, ranging from a rapidly spinning BH with a retrograde accretion flow (left) to a non-spinning BH (center) to a rapidly spinning BH with a prograde accretion flow (right). The angular radius of the black hole, $M/D$, is identical in each panel. 
    The polarization pattern becomes more radial at higher spin, as frame dragging enforces toroidal magnetic fields near the horizon. In retrograde flows, the spirals pattern reverses handedeness over radius, indicating the transition from the prograde rotation within the ergosphere to the retrograde flow at larger radii. The handedness flips across sub-images, leading to depolarization in the photon ring of the full image \citep[see][]{Jimenez_Rosales_2021,Palumbo_Wong_2022}. By studying the polarized structure and its radial evolution, the ngEHT can estimate the spin of \m87 and \sgra and quantify the effects of frame dragging. 
    Adapted from \citet{Palumbo_2023}.
    }
    \label{fig:spin_pol}
\end{figure*}

The ngEHT has the opportunity to provide decisive measurements of spin through several approaches \citep[for a summary of these methods, see][]{Ricarte_2022_spin}. The most compelling method would be to study the detailed structure of the lensing signatures such as the photon ring (see \autoref{sec:Photon_Ring}), or the (inner) shadow (see \autoref{sec:Horizon_Physics}). However, while spin has a pronounced effect on these features, the effects of spin manifest on scales that are still much smaller than the nominal resolution of the ngEHT, so a conclusive detection may not be possible. Nevertheless, the effects of spin may be apparent in the emission structure on somewhat larger scales, particularly through the polarization structure in the emission ring \citep[see \autoref{fig:spin_pol} and][]{Palumbo_2020,Qiu_2022}. Finally, spin signatures are expected to be imprinted in the time-domain.

At least initially, ngEHT estimates of spin will likely rely on numerical simulations because unambiguous signatures of spin would require significantly finer angular resolution. These estimates will likely require confirmation through multiple lines of study---total intensity, polarization, and time-domain---and through a variety of modeling approaches including semi-analytic studies \citep[e.g.,][]{Palumbo_2022}. Current studies indicate that the time-averaged polarized structure of \m87 is the most reliable estimator of spin, with 345\,GHz observations essential to improving angular resolution and also to quantify the potential effects of internal Faraday rotation on the polarized structure \citep[see, e.g.,][]{Moscibrodzka_2017,Ricarte_2020,Ricarte_2023_pol}.

\subsubsection{Constraining the Properties of a Black Hole's Photon Ring}
\label{sec:Photon_Ring}

The image of a BH is determined by two different factors: the complex astrophysical phenomena in its vicinity, which are the source of the emergent electromagnetic radiation, and the spacetime geometry, which introduces effects such as gravitational lensing and redshift. To isolate relativistic effects requires disentangling the complex, turbulent astrophysical environment from the comparatively simple spacetime dependence. 
Gravitational lensing is particularly useful in this context, as it gives rise to matter-independent (``universal'') features, such as the ``photon ring.'' The photon ring is a brightness enhancement along an approximately circular closed curve on the image, which arises from light rays undergoing multiple half-orbits around the BH before reaching the telescope \citep{Johnson_2020}. These rays are small deviations from the unstable bound spherical orbits near a Kerr BH \citep{Bardeen_1973,Teo_2003}. 
We index these half-orbits with the number $n$; the observer sees exponentially demagnified images of the accretion flow with each successive $n$ (see \autoref{fig:PhotonRingStack}). At the resolution of Earth baselines at 230\,GHz and 345\,GHz, only $n = 0$ and $n = 1$ emission is likely to be detectable. Because the ngEHT cannot resolve the thickness of the primary ($n=0$) ring, ngEHT studies of the photon ring necessarily require some degree of super-resolution, with associated model-dependent assumptions. In general, the principal challenge for ngEHT studies of the photon ring is to unambiguously disentangle the signals of the primary and secondary photon rings \citep[see, e.g.,][]{Tiede_2022}. 

In the asymptotic ($n \rightarrow \infty$) limit, the photon ring has an intricate and universal structure which depends only on the spacetime geometry and acts as a lens for electromagnetic radiation \citep[e.g.,][]{Takahashi_2004,Johannsen_Psaltis_2010}. However, even at small-$n$, the photon ring carries information on the BH's mass and spin and provides a novel strong-field test of general relativity \citep{Wielgus2021,Broderick_relative_rings}, especially if combined with a strong independent mass measurement (e.g., as is given by resolved stellar orbits of \sgra; see \autoref{fig:EHT_SgrA}). A clear goal for the ngEHT is to use the improved angular resolution and sensitivity to constrain the properties of the photon rings in \m87 and \sgra.

Tests with both geometric model fitting of the sky intensity distribution and emissivity modelling in the BH spacetime suggest that the long baselines at 345\,GHz are a strict requirement for detecting the $n = 1$ ring \citep{Palumbo_2022,Tiede_2022}. While intermediate baselines are required to support these model-fitting approaches, achieving the highest possible angular resolution is the driving requirement for studies of the photon ring. Photon ring detection using time-averaged images is likely most relevant to \m87, as \sgra observations are expected to be severely affected by scattering in the ionized interstellar medium \citep{Psaltis_2018,Johnson_2018,Issaoun_2019,Zhu_2019}. Alternatively, signatures of the photon ring may be accessible in the time-domain, where ``light echoes'' can appear from either impulsive events such as flaring ``hot spots'' or from stochastic fluctuations in the accretion flow \citep[see, e.g.,][]{Broderick_2006,Moriyama_2019,Tiede_2020, Chesler2021,Hadar_2021,Wong_2021,Gelles_2021,Gelles_2022,Wielgus_2022,Vos2022,Emami_2022_hotspot}.

\subsubsection{Constraining Ultralight Fields}
\label{sec:axions}

The existence of ultralight boson fields with masses below the eV scale has been predicted by a plethora of beyond-Standard-Model theories \citep[e.g.,][]{Peccei_Quinn_1977,Preskill_1983,Abbott_Sikivie_1983,Dine_Fischler_1983,Arvanitaki:2009fg}. Such particles are compelling dark matter candidates and are, in general, extremely hard to detect or exclude with usual particle detectors. However, quite remarkably, rotating BHs can become unstable against the production of light bosonic particles through a process known as BH superradiance \citep{Brito:2015oca}. This process drives an exponential growth of the field in the BH exterior, while spinning the BH down. Superradiance is most effective for highly spinning BHs and when the boson's Compton wavelength is comparable to the BH’s gravitational radius \citep{Brito:2015oca}. A BH of mass ${\sim}10^{10} M_\odot$ such as \m87 can be superradiantly unstable for ultralight bosons of masses $10^{-21}\,{\rm eV}$ \citep[this particular value leads to ``fuzzy'' dark matter, predicting a flat distribution that is favored by some observations;][]{Hu:2000ke}.

For very weakly interacting particles, the process depends primarily on the mass and spin of the BH, and on the mass and spin of the fundamental boson. By requiring the predicted instability timescale to be smaller than the typical accretion timescale (that tends to spin up the BH instead), one can then draw regions in the parameter space where highly spinning BHs should not reside, if bosons within the appropriate mass range exists in nature. Thus, BH spin measurement can be used to constrain the existence of ultralight bosons. In particular, obtaining a lower limit on the BH spin is enough to place some constraints on boson masses (with the specific boson mass range constraint dependent on the BH spin). This approach is practically the only means to constrain weakly interacting fundamental fields in this mass range. \citet{Davoudiasl_axion_2019} used this line of argument to constrain masses of ultralight boson dark matter candidates with the initially reported EHT measurements that the SMBH must be spinning to produce sufficient jet power \citepalias{EHTC_M87_V}.

\begin{figure*}[t]
\centering
\includegraphics[width=\textwidth]{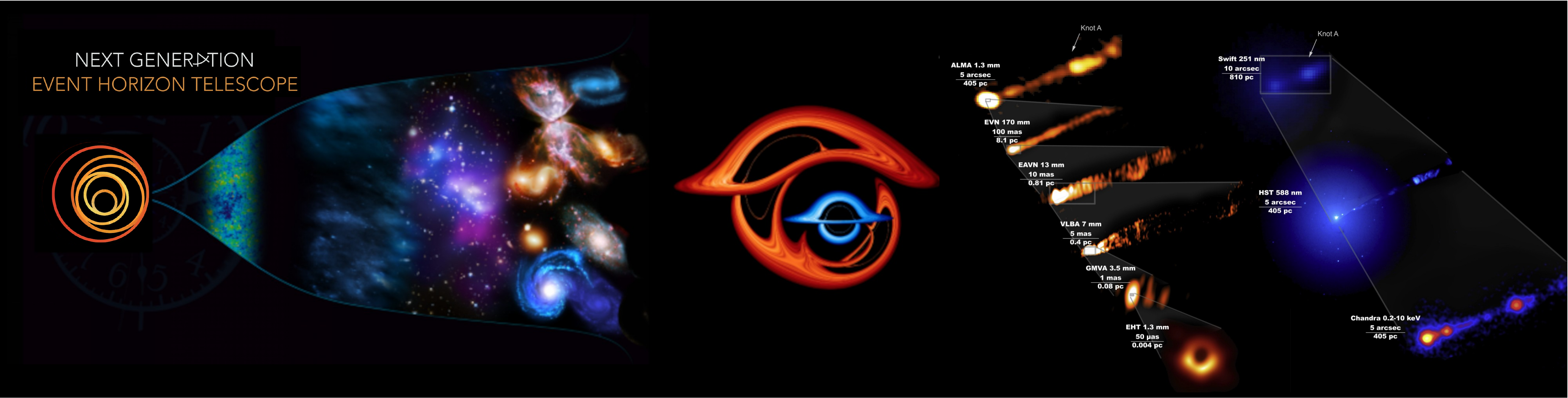}
\caption{Conceptual illustration of the science cases explored within the ``Black holes and their cosmic context'' SWG: BH growth, binary BHs and gravitational waves, and MWL studies of BHs and jets. Credits from left to right: Perimeter Institute, NASA's Goddard Space Flight Center/Jeremy Schnittman and Brian P. Powell, J. C. Algaba for the EHT Collaboration \citep{EHT_MWL}. Composition: Thalia Traianou, IAA-CSIC.}
\label{fig:BHCC}
\end{figure*}

Among all the families of suggested ultralight particles, axions are one of the best studied and most highly motivated from a particle physics perspective. For axions with strong self-interactions, the super-radiance process will end up with a weakly saturating phase where the axion field saturates the highest possible density in the Universe. Due to the axion-photon coupling, the coherently oscillating axion field that forms around the BH due to superradiance can give rise to periodic rotation of the electric vector position angle (EVPA) of the linearly polarized emission. The amplitude of the EVPA oscillation is proportional to the axion-photon coupling constant and is independent of the photon frequency. The variations of the EVPA behave as a propagating wave along the photon ring for a nearly face-on BH. For instance, using the 4 days of polarimetric measurements of \m87 published by the EHT collaboration in 2021 \citepalias{EHTC_M87_VII}, one can already constrain the axion-photon coupling to previously unexplored regions \citep{Chen_2020,Chen_2022}. The upper bound on the axion mass window is determined by the spin of the BH via the condition for superradiance to occur.

For improved constraints on these fundamental fields and their electromagnetic couplings, the ngEHT must observe polarimetric images of \m87 in a series of at least 3 days over a 20-day window (the expected oscillation period). As for other cases that rely on polarimetry, observations at both 230 and 345\,GHz are imperative to isolate the potential effects of Faraday rotation, and repeated observations will be needed to distinguish periodic oscillations from stochastic variability \citep{Chen:2022oad}.

\subsection{Black Holes and their Cosmic Context}
\label{sec:BHCC}

The growth of SMBHs is driven primarily by gas accretion and BH-BH mergers. Mergers are expected to dominate low-redshift SMBH growth in dense environments, especially in the high mass range to which the ngEHT will be most sensitive \citep{Kulier+2015, Weinberger+2018,Ricarte&Natarajan2018a,Pacucci&Loeb2020}.  Gas accretion onto SMBHs is a critical piece of the current galaxy formation paradigm, in which feedback from accreting SMBHs is required to regulate gas cooling and star formation in massive galaxies \citep[e.g.,][]{HaehneltPNRees1998, DiMatteo+2005,Croton+2006}.  At present, however, the details of the feedback processes are poorly understood and are currently the largest source of uncertainty in understanding the combined mass assembly history and evolution of galaxies and their central SMBHs.

The ngEHT will provide unique observational access to both modes of SMBH growth through studies that extend over a vast range of scales (see \autoref{fig:BHCC}).  By beginning to resolve the accretion flows of dozens of AGNs, the ngEHT will enable the extraction of information on their masses, spins, and accretion rates, providing crucial insights into their mass assembly history and growth (\autoref{sec:BH_Formation_Growth}).  In addition, the ngEHT will have sufficient angular resolution to identify sub-parsec binary SMBHs at any redshift, providing a powerful complement to gravitational wave observations of galaxy mergers (\autoref{sec:binaries}). In addition, the ngEHT will provide new insights into how SMBHs influence their galactic environments via feedback through multi-wavelength and multi-messenger studies of their relativistic outflows. (\autoref{sec:mwl_mm}). We now discuss the goals and requirements associated with each of these objectives.

\subsubsection{Understanding Black Hole-Galaxy Formation, Growth and Coevolution}
\label{sec:BH_Formation_Growth}

The masses and spins of SMBHs encode their assembly history.  SMBH masses trace this assembly history in a statistical fashion, with the distribution of SMBH masses -- i.e., the BH mass function (BHMF) -- capturing the population-level growth and evolution over cosmic time \citep{Kelly_2012}.  Measurements of SMBH spin can trace the growth histories of individual objects.  For instance, BHs accreting from a thin disk with a steady rotation axis can be spun up to a maximum value of $a=0.998$ \citep{Thorne1974}, while discrete accretion episodes from disks with random rotation axes will tend to spin a BH down \citep{King+2008}.  In addition, BHs accreting at low-Eddington rates for Gyrs can also be spun down due from the energy extraction that is required to power their jets via the Blandford-Znajek process \citep{Blandford_Znajek_1977,Narayan+2021}.

\begin{figure*}[t]
\centering
\includegraphics[width=0.49\textwidth]{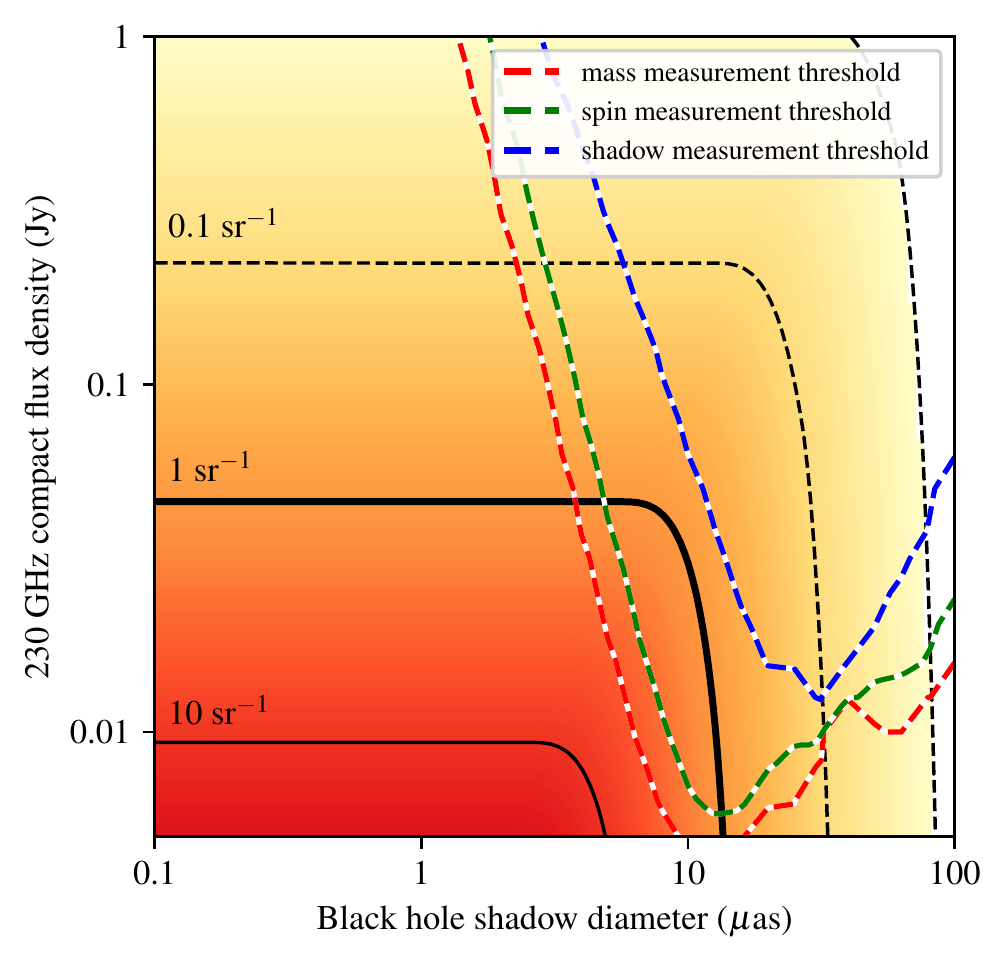}\hfill
\includegraphics[width=0.49\textwidth]{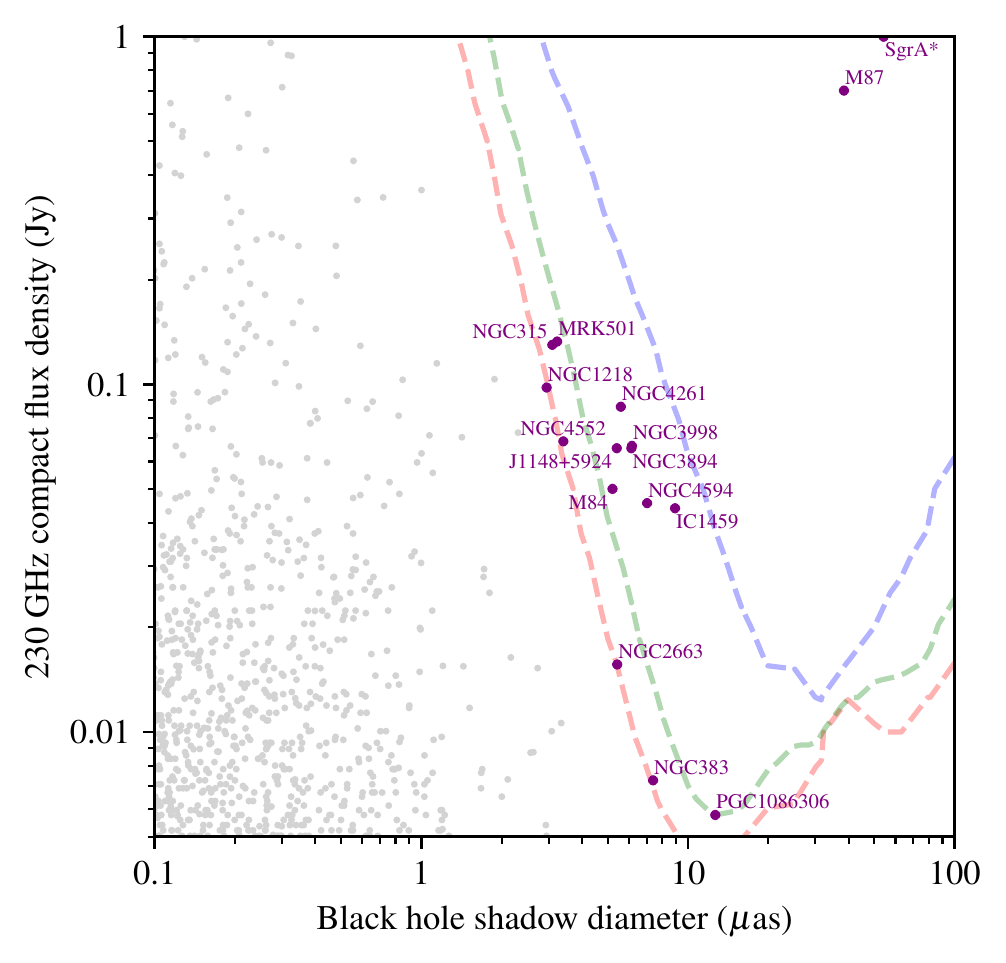}
\caption{
SMBH population studies with the ngEHT. 
(left) Black contours show the estimated cumulative number density of SMBHs as a function of shadow diameter and 230\,GHz flux density. Colored contours indicate threshold values at which the ngEHT Phase-1 could plausibly measure the SMBH mass (red), spin (green), and shadow (blue) in a superresolution regime. Reproduced from \citet{Pesce_2022}. 
(right) Estimated 230\,GHz compact flux density and BH shadow diameter for a subset of bright VLBI-detected SMBHs in the ETHER database. Colored lines again indicate the approximate measurement thresholds for the ngEHT Phase-1 array to measure the BH mass, spin, and shadow as shown on the left. Adapted from \citet{Ramakrishnan_2023}. 
}
\label{fig:Demographics}
\end{figure*}

The ngEHT will provide access to SMBH masses by observing the sizes of their horizon-scale emitting regions at (sub)millimeter wavelengths.  EHT observations of \m87 have demonstrated that measurements of the diameter of the ring-like emission structure can be used to constrain the SMBH mass \citepalias{EHTC_M87_VI}.  The $\sim$11\% mass measurement precision achieved using the initial EHT observations of \m87 -- and even the comparatively modest $\sim$25\% precision achieved for the more challenging observations of \sgra \citepalias{EHTC_SgrA_IV} -- establish the ``shadow size technique'' as among the most precise means of measuring SMBH masses \citep[see, e.g.,][]{Kormendy_2013}.  With the additional angular resolution and sensitivity provided by the ngEHT, \citet{Pesce_2022} estimate that $\sim$50 SMBH masses will be measurable for nearby AGNs distributed throughout the sky (see \autoref{fig:Demographics}).  These measurements will substantially increase the number of SMBHs with precisely-measured masses, improving our understanding of the BHMF in the local Universe.

Relative to mass measurements, observational spin measurements for SMBHs are currently scarce; only roughly three dozen spin measurements are available for nearby SMBHs, with the majority obtained from X-ray diagnostics of the iron K-alpha line \citep{Reynolds_2021}.  These iron-line measurements are uncertain because the method is highly sensitive to the orbital radius at which the accretion disk's inner edge truncates, which is typically assumed to occur at the innermost stable circular orbit \citep{Brenneman2013}.  In addition to their large uncertainties, current X-ray measurements are also biased towards high Eddington ratio objects.

The ngEHT will provide access to SMBH spins by observing the polarized radio emission emitted by the horizon-scale accretion flows around nearby AGNs.  Current EHT observations have provided only modest constraints on the spin of \m87 \citepalias{EHTC_M87_V,EHTC_M87_VIII}, but recent and ongoing advances in our theoretical understanding of near-horizon accretion flows will soon enable more precise spin quantifications from similar observations.  As detailed in \citet{Ricarte_2022_spin}, linear polarimetric observations made by the ngEHT will provide estimates of SMBH spins by tracing the near-horizon magnetic field structures.  The curl of the linear polarization pattern in the emission ring near a SMBH has been shown to correlate with SMBH spin in GRMHD simulations \citep{Palumbo_2020,Qiu_2022}.  Ongoing studies indicate that this correlation originates from changes in the magnetic field geometry that are associated with frame dragging, which becomes stronger as spin increases \citep{Emami_2022_pol}.  \citet{Pesce_2022} estimate that the ngEHT will be able to constrain $\sim$30 SMBH spins through measurements of their horizon-scale polarized radio emission.  Moreover, the spin measurements enabled by the ngEHT will offer fundamentally new insights by constraining the spins of low Eddington ratio SMBHs -- rather than the high Eddington ratio SMBHs preferentially measured using X-ray techniques -- which is a regime that is more representative of the overall SMBH population in the Universe.

The estimates from \citet{Pesce_2021} and \citet{Pesce_2022} for the number of SMBHs for which the ngEHT can make mass and/or spin measurements are based on statistical considerations, using our current understanding of the local BHMF and the distribution of SMBH accretion rates to predict how many objects should fall within the observable window.  However, identifying the specific objects to target with the ngEHT for these measurements requires dedicated observational surveys of AGN to determine which sources are sufficiently bright, massive, and nearby.  To this end, the Event Horizon and Environs \citep[ETHER; ][]{Ramakrishnan_2023} database aims to provide a standardized catalog of ngEHT targets.  Currently, the ETHER sample includes ${\sim}10^3$ SMBHs that have been previously observed to have mas-scale structure at cm wavelengths and which have predicted 230\,GHz flux densities greater than a few mJy.  Of these sources, $\sim$10 have bright 8--86 GHz VLBI detections (from jet emission) and are predicted to be bright enough to image their jet bases at $\lesssim$100 R$_{\rm g}$ with the ngEHT (see \autoref{fig:Demographics}).  The identification of ngEHT targets with bright accretion inflows but without detected cm-wave jets is ongoing; the currently known $\sim$200 BHs with estimated ring sizes ${\geq}5$\,$\mu$as primarily have (observed arcsec-scale and/or predicted mas-scale) 230\,GHz flux densities less than 1\,mJy, with the brightest falling in the 1 to 10\,mJy range.  The upcoming release of the e-ROSITA all-sky hard X-ray survey (with SDSS V and 4MOST spectroscopic followups for BH mass estimates) is expected to significantly expand the list of potential targets in this accretion-inflow-only sample, which will permit definitive specifications on the sensitivity requirement for the ngEHT to measure a large population of horizon-resolved sources.

\subsubsection{Understanding how SMBHs merge through Resolved Observations of Sub-Parsec Binaries }
\label{sec:binaries}

Binary SMBHs are generic products of galaxy mergers, that are thought to drive structure formation in our dark energy driven cold dark matter Universe. During SMBH mergers, dynamical friction and stellar mass segregation act to draw the two resident massive objects to the center of the merger remnant \citep{Merritt_2005}. The environmental interactions that drive the binary to separations of ${\sim}0.1$--10\,pc are understood, but the mechanism(s) that drive continued inspiral beyond this point -- and in particular, to the sub-parsec regime in which gravitational wave emission is expected to efficiently complete the merger process -- remain unclear \citep[e.g.,][]{Begelman_1980}.  A number of solutions to this long-standing and so-called ``final parsec problem'' \citep{Armitage_2002,Milosavljevic_2003} have been proposed; for instance, interactions with gas in a circumbinary disk, and three-body interactions with stars could all contribute and have significant influence on the shape and evolutionary timescale of the binary.  Uncovering the details of the physics in this last parsec informs the science cases of future gravitational-wave detectors such as Pulsar Timing Arrays (PTAs) and space-based gravitational-wave interferometry (e.g., LISA).

The ngEHT will have a nominal angular resolution of 15\,$\mu$as, which implies a linear resolution of $\leq$0.13\,pc across all redshifts.  The effective resolving power may be further improved by a factor of several through the use of ``super-resolution'' techniques (e.g., \citealt{Chael_2016}, \citealt{Akiyama_2017a}, \citealt{Broderick_hybrid}). The ngEHT can therefore \textit{spatially resolve} SMBH binaries that have entered their steady-state gravitational wave emission phase.  The orbital period at this stage is typically short (months to years), which makes it accessible to multi-epoch observations with the ngEHT.  Furthermore, \citet{DOrazio_2018} estimate that between $\sim$1--30 sub-parsec SMBH binaries should have millimeter flux densities in the $\gtrsim$1\,mJy regime that will be accessible with the ngEHT.

\subsubsection{Multi-wavelength and Multi-Messenger Studies of SMBHs and their Relativistic Outflows}
\label{sec:mwl_mm}

The EHT has already demonstrated the immense value of extensive multi-wavelength campaigns to augment horizon-scale imaging \citep[e.g.,][]{EHT_MWL,EHTC_SgrA_II}, and the ngEHT will similarly benefit from coordinated observations \citep[for a review, see][]{Lico_2023}. In particular, the relativistic jets launched by SMBHs extend the gravitational influence of BHs to galactic scales, converting and transporting immense amounts of energy across the full electromagnetic spectrum. These jets act as powerful particle accelerators that are thought to produce ultra-high energy cosmic rays and have also been implicated in the production of high-energy neutrinos \citep[e.g.,][]{IceCube_2018,IceCube_2018b,Plavin_2020,Plavin_2021,Giommi_2021,Plavin_2023}. 

The ngEHT can directly image flaring regions, creating an opportunity to shed light on the physical mechanisms that drive acceleration of protons to PeV energies and generation of high-energy neutrinos. 
Moreover, crucial insights into the jet composition can be obtained by combining information about the jet dynamics with information about the accretion power (e.g., from x-ray observations).  
Ideally, this will involve both triggered and monitoring ngEHT observations. Triggered observations would happen when a neutrino arrives within an error region from a strong blazar. Limiting the trigger on both the neutrino energy (above ${\sim}100\,{\rm GeV}$) and VLBI flux density (above ${\sim}0.5\,{\rm Jy}$) would increase the probability of association and ensure sufficiently high dynamic range of the ngEHT images, respectively. The initial trigger would be followed by ${\sim}$monthly monitoring for a year. In addition to this mode, observing a large sample of the strongest blazars with ${\sim}$monthly monitoring, supplemented by additional single-dish flux monitoring, will provide an opportunity to study the evolution of sources before neutrino production and to characterize the features that are associated with neutrino production.

In addition, by observing a population of blazars, the ngEHT will be able to measure the jet profile from the immediate vicinity of a SMBH through the acceleration and collimation zones and past the Bondi radius \citep[e.g.,][]{Kovalev_2020}. By observing with coordinated multi-wavelength campaigns, the ngEHT will provide decisive insights into the nature of the bright, compact ``core'' feature that is seen in many blazars \citep[e.g.,][]{Marscher_2008}. Current EHT images of blazars show complex, multi-component emission \citep{Kim_2020,Issaoun_2022,Jorstad_2023}, 
%\citepalias[e.g., \citealt{Kim_2020};][]{Issaoun_2022}, 
so ngEHT observations extending over multiple months to study the evolution of these components will be imperative. 

Multi-wavelength and multi-messenger studies of flaring activity in blazar jets will require ngEHT monitoring campaigns with triggering capabilities followed by a cadence of the order of weeks. Full Stokes polarization capabilities with high accuracy (systematic errors on polarization ${\lsim}\,0.1\%$) and high imaging dynamic range (${\gsim}\,1000{:}1$ to detect faint jet emission) will be required for mapping the magnetic field in the jet regions through Faraday rotation analyses. Close coordination with other next-generation instruments, such as the Cherenkov Telescope Array (CTA), LISA, SKA, ngVLA, and Athena will significantly enrich the potential for multi-wavelength and multi-messenger studies with the ngEHT.

%%%%%%%%%%%%%%%%%%%%%%%%%
\subsection{Accretion}
\label{sec:Accretion}
%%%%%%%%%%%%%%%%%%%%%%%%
Electromagnetic radiation from SMBHs such as \m87 and \sgra originates in hot gas, which is brought close to the BH by an accretion disk \citep[for a review of hot accretion flows, see][]{Yuan_Narayan_2014}. Some of the same gas is also expelled in relativistic jets or winds. Spatially resolved images of the disk and its associated dynamics provide a remarkable new opportunity to study accretion physics. 

BH accretion disks are believed to operate with the help of the magnetorotational instability,\footnote{Angular momentum transport may also occur in magnetic flux eruptions \citep[see, e.g.,][]{Chatterjee_2022}, which would also have distinctive signatures in ngEHT images and movies \citep[see, e.g.,][]{Ripperda_2022,Gelles_2022,Jia_2023}.} which amplifies the magnetic field in the plasma and uses the associated shear stress to transport angular momentum outward \citep{Balbus_Hawley_1991,Balbus_Hawley_1998}. Signatures of the magnetic field are revealed via linear and circular polarization of the emitted radiation. Yet, while spatially-resolved and time-resolved spectropolarimetric observations are thus exceptional tools for studying the inner workings of BH accretion, \emph{we do not at present have even a single spatially-resolved image of any BH accretion disk.}

The closest current results are through EHT observations of \m87 and \sgra. The ring-shaped 230\,GHz emission surrounding a central brightness depression confirms strong light deflection and capture near these BHs. However, the angular resolution and dynamic range achieved so far by the EHT are modest, and it is unclear what part of the observed radiation is from the accretion disk and what is from the jet \citepalias[see, e.g.,][]{EHTC_M87_V,EHTC_SgrA_V}. The ngEHT will have the sensitivity to image out to larger radii from the BH and to make time-resolved movies in all Stokes parameters. These advances will enable progress on three broad fronts in accretion physics: revealing the physical mechanism that drives accretion onto SMBHs (\autoref{sec:accretion_driver}), observing localized electron heating and dissipation (\autoref{sec:accretion_flares}), and measuring signatures of frame dragging near a rotating black hole (\autoref{sec:frame_dragging}).

\subsubsection{Revealing the Driver of Black Hole Accretion}
\label{sec:accretion_driver}

Our current understanding of accretion close to a BH is largely guided by ideal general relativistic magnetohydrodynamical (GRMHD) numerical simulations \citep[see, e.g.,][]{Gammie_2003,Porth_2019}. These simulations suggest that the strength and topology of the magnetic field play an important role. When the field is weak and scrambled, the accreting gas becomes turbulent, with eddies over a wide range of length scales \citep[e.g.,][]{Narayan_2012}. When the field is strong, and especially when it also has a dipolar configuration \citep[this is called a ``magnetically arrested disk'' or MAD;][]{Narayan_2003}, accretion occurs via large discrete inflowing streams punctuated by episodic outward eruptions of magnetic flux. The ngEHT will be able to identify these and other dynamical patterns in the accretion flow by making real-time movies. Flux-tube eruptions \citep{Chan_2015,Ripperda_2022,Chatterjee_2022,Gelles_2022}, orbiting spiral patterns \citep[e.g.,][]{Guan_2009}, and bubbling turbulence, could all be accessible to observations. Crucially, spatially-resolved measurements of the linear polarization fraction, degree of circular polarization, and Faraday rotation, will provide rich detail on the magnetic field topology and its strength \citep[e.g.,][]{Ricarte_2023_pol}. Different target sources will presumably have different dynamics and field configurations, opening up a fruitful area of research. In the specific case of a MAD system, it is unknown exactly how the strong field originates. One proposal posits that the field is generated in situ by a radiation-driven battery mechanism \citep[e.g.,][]{Contopoulos_2018}. It predicts a specific relative orientation of the dipolar magnetic field with respect to the accretion disk angular velocity vector. If any of ngEHT's targets is MAD (EHT observations suggest \m87 and \sgra may both be such systems), testing the predictions of the radiation battery model would be an important secondary goal.

Accretion-related ngEHT science will be primarily enabled through observations of \m87 and \sgra, with two major associated challenges. First, the most interesting effects occur in regions of the disk within a few event horizon radii. But this is precisely where the observed image is highly distorted by the gravitational lensing action of the BH, the same effect which produces the ring image of \m87. Disentangling lensing to reveal the true underlying structure of the accretion disk will require new image processing techniques. Second, the observed image will often be a superposition of radiation from the accretion disk and the jet. The two components will need to be separated. One promising method is to utilize dynamics and variability, which can be quite different in the disk and in the jet. Observations with a cadence of $t_{\rm g}$ would be ideal to study the most rapid variability, and interesting variations are expected on all timescales up to $10^3 - 10^4\, t_{\rm g}$. Full-night observations with sufficient baseline coverage for snapshot imaging on sub-minute timescales will needed for \sgra, while a monitoring campaign with a sub-week cadence and extending for at least 3 months (and, ideally, over multiple years) will be ideal for \m87.

\subsubsection{Localized Heating and Acceleration of Relativistic Electrons}
\label{sec:accretion_flares}

The radiation emitted from an accretion disk is produced by hot electrons, which receive their heat energy via poorly-understood plasma processes in the magnetized gas. The most promising idea for heating is magnetic reconnection, which can occur in regions with large-scale topological reversals of the magnetic field, or in regions with large shear, or where small-scale turbulent eddies dissipate their energy. All of these processes are at their most extreme in the relativistic environment found in BH accretion disks.

Our current understanding of relativistic magnetic reconnection is based on particle-in-cell (PIC) simulations \citep[e.g.,][]{Sironi_2014,Rowan_2017,Werner_2018,Ball_2018}. These numerical studies show clear evidence for unequal heating of electrons and ions, as well as acceleration of both into a non-thermal distribution with a power-law tail at high energies. Electron heating in large flares in BH disks would be especially interesting for ngEHT observations. A flare may initially appear as a bright localized region in the image. It will subsequently move around the image, will also likely spread to become more diffuse, and will show effects from graviational lensing \citep[see, e.g.,][]{Broderick_2006,Doeleman_2009,Tiede_2020,Gelles_2022}. Both the ordered motion of the heated region and its spreading will provide fundamental information on the microscopic plasma physics processes. The heated electrons will also cool as they radiate, causing the electron distribution function (eDF) to evolve. Multi-wavelength imaging will provide a handle on both the dynamics and the eDF evolution.

Less dramatic steady heating should also be present, and it will likely show strong variations as a function of radius in both amplitude and eDF. With the enhanced dynamic range of the ngEHT, these spatial variations should be accessible over a factor of 10 range of radius. Particle acceleration and heating is relevant for a wide range of astrophysical phenomena. While we have some information on low energy processes from laboratory experiments and measurements in the solar wind, there is currently no observational technique for direct study of heating in relativistic settings. Imaging BH accretion disks with the ngEHT can reveal localized heating and acceleration on astrophysical scales and will track the evolution of the energized electrons. Lessons from such observations would have widespread impact in many other areas of astrophysics.

%%%%%%%%%%%%%%%%%%%%%%%%%%%%%%%%%%%%%%%%%%%%%%%%%%%%%%%%%%%%%%%%%%%%%
\subsubsection{Dynamical Signatures of Frame Dragging near a Rotating Black Hole}
\label{sec:frame_dragging}
%%%%%%%%%%%%%%%%%%%%%%%%%%%%%%%%%%%%%%%%%%%%%%%%%%%%%%%%%%%%%%%%%%%%%
Direct observations of the inner region of the accretion disk provide an opportunity to study the object at the center, namely, the BH itself. While the most significant effect on large scales is the immense gravitational pull of a BH, another gravitational property of these objects is arguably even more interesting. Namely, a spinning BH has the remarkable property that it drags space around it in the direction of its spin. This so-called frame-dragging effect is felt by all objects outside the BH, including the accretion disk. The effect is strongest in regions within a few event horizon radii of the BH. 

Spatially-resolved and time-resolved imaging has the potential to confirm the frame-dragging effect and to study its details \citep[see, e.g.,][]{Ricarte_2022_frame_dragging}. Since the accretion disk is fed by gas at a large distance from the BH, the outer disk's angular momentum vector is likely to be randomly oriented with respect to the BH spin axis. Only when gas comes close to the BH does it feel the spin direction of the BH via frame-dragging. The manner in which the disk adjusts its orientation can provide direct confirmation of the frame-dragging phenomenon. If the disk is tilted with respect to the BH spin vector, it is expected to precess and align with the BH inside a certain radius \citepalias[see, e.g.,][]{EHTC_M87_V}. Both the precession and alignment can be observed and studied by the ngEHT. In the special case of a retrograde accretion flow (i.e., when the disk's orbital motion is in the opposite direction to the BH spin), the angular velocity of the disk gas will reverse direction close to the horizon. There will be a related effect also in the orientation of the projected magnetic field, which may be visible in polarimetric ngEHT images. Observing these effects directly with the ngEHT would be a breakthrough achievement and would provide a new tool to study a central prediction of the Kerr spacetime (see also \autoref{sec:Spin}).

\begin{figure*}[t]
    \centering
    \includegraphics[width=\textwidth]{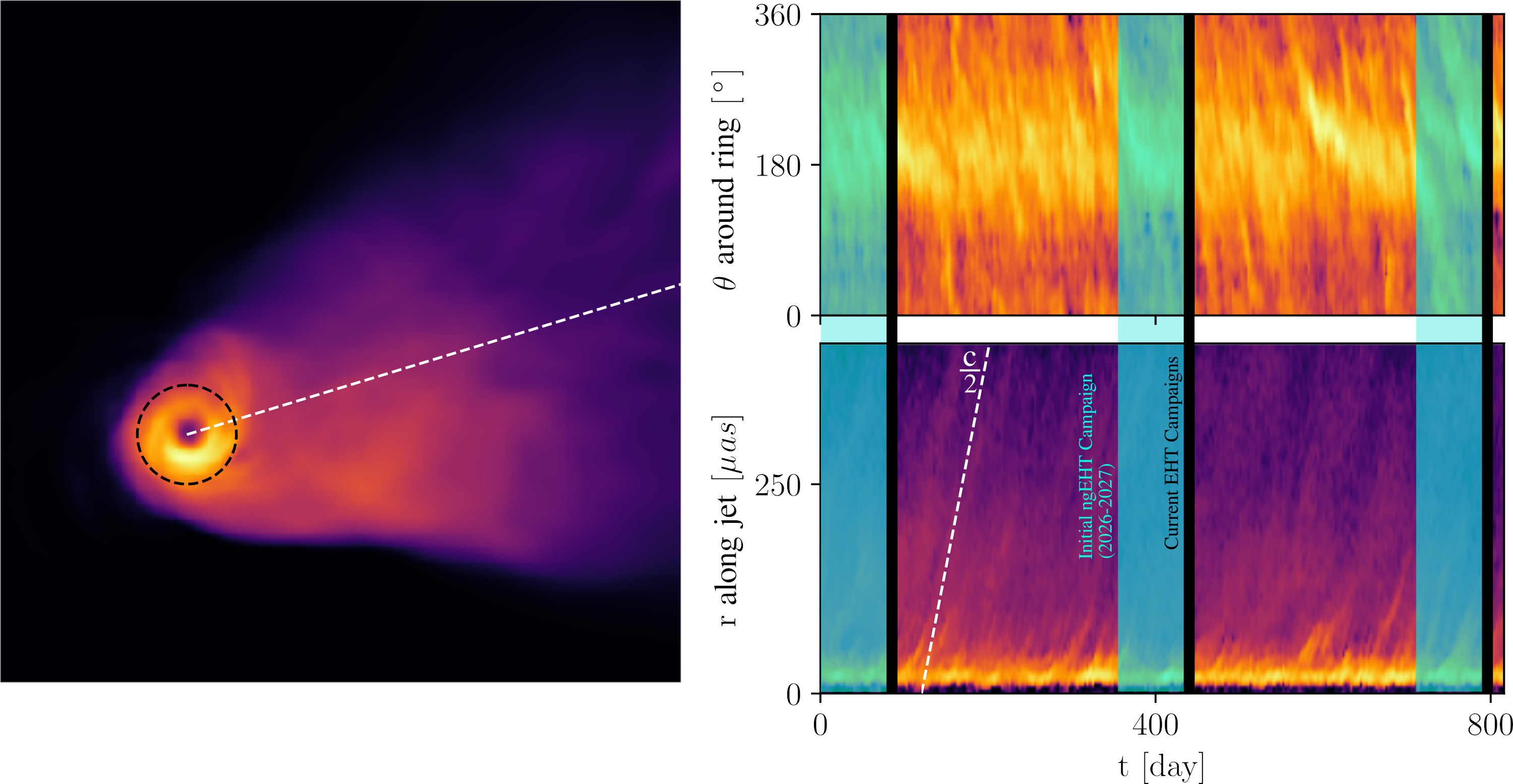}
    \caption{Studying accretion and jet dynamics with the ngEHT. (left) A frame from a simulated movie of \m87 \citep{Chael_2019}. (right) Azimuthal (top) and radial (bottom) brightness variations in a reconstructed movie of \m87 using ngEHT Phase-1 coverage. The top panel shows how azimuthal variations around the black dashed circle track orbital dynamics near the BH, evident here as diagonal striations with sub-Keplerian angular velocity. The bottom panel shows how radial variations along the white dashed line will reveal the SMBH-jet connection and measure acceleration within the jet-launching region. Initial ngEHT monitoring campaigns (light blue vertical bands) will span 3~months per year with a dense (sub-week) observing cadence; for comparison, current EHT campaigns (dark vertical bands) only span ${\sim}2\,{\rm weeks}$ per year, which is insufficient to measure the dynamics of the accretion disk or jet.
    }
\label{fig:Jet_Dynamics}
\end{figure*}

\subsection{Jet Launching}
\label{sec:Jet_Launching}

Relativistic jets are among the most energetic phenomena in our universe, emitting radiation throughout the entire electromagnetic spectrum from radio wavelengths to the gamma-ray regime, and even accelerating particles to highest measured energies \citep[for a review, see][]{Blandford_2019}. The most powerful jets are those that are anchored by nuclear SMBHs in AGN, as emphatically demonstrated through the images of \m87 with the EHT. Yet, despite this impressive breakthrough, the actual jet launching mechanism and power source is still uncertain. The ngEHT has the potential to make pivotal discoveries related to the power source of relativistic jets (\autoref{sec:jet_power}), and to the physical conditions that launch, collimate, and accelerate these jets (\autoref{sec:Jet_Conditions}).

%%%%%%%%%%%%%%%%%%%%%%%%%%%%%%%%%%%%%%%%%%%%%%%%%%%%%%%%%%%%%
\subsubsection{Jet Power and Black Hole Energy Extraction}
\label{sec:jet_power}
%%%%%%%%%%%%%%%%%%%%%%%%%%%%%%%%%%%%%%%%%%%%%%%%%%%%%%%%%%%%
According to our current theories, jets can either be powered by the liberation of gravitational potential energy in the accreting material \citep[e.g.,][]{Salpeter_1964} or by directly extracting the rotational energy of a spinning BH \citep{Blandford_Znajek_1977}. In both processes, magnetic fields must play a crucial role. Therefore, measuring the velocity field of the innermost jet regions and comparing de-projected rotation of the magnetic fields with the rotation of the BH ergosphere will probe whether jets are launched by rotating BHs. 

The ideal target to address this question is \m87 because of its large BH mass ($M \approx 6.5 \times 10^9\,M_\odot$), proximity ($D \approx 16.8\,{\rm Mpc}$), and prominent jet ($P_{\rm jet} \gg 10^{42}\,{\rm erg/s}$).  \sgra also provides an important target to study---despite decades of VLBI observations, there is no firm evidence for a jet in \sgra at any wavelength. Nevertheless, there are compelling reasons to continue the search for a jet in \sgra with the ngEHT, including the potential for interstellar scattering to obscure the jet at longer wavelengths \cite[e.g.,][]{Markoff_2007}, evidence for an outflow in frequency-dependent time lags during flares \citep[e.g.,][]{YusefZadeh_2006,Brinkerink_2021}, and the fact that favored GRMHD models for \sgra based on constraints from EHT observations predict the presence of an efficient jet outflow \citepalias{EHTC_SgrA_V}. Comparing the jets in \m87 and \sgra, together with knowledge of their respective BH properties, will provide fundamental insights into the role of the BH and its environment in producing a jet.

Current EHT observations are limited both in terms of the baseline coverage and image dynamic range, which prohibits estimates of physical parameters in the critical region just downstream of the BH. The ngEHT will provide superior baseline coverage and increased dynamic range, allowing reconstructed movies that simultaneously resolve horizon scale structure and the jet base in \m87 and \sgra. To identify the source of the jet's power with the ngEHT will require estimates of the magnetic flux threading the SMBH, the spin of the SMBH (see \autoref{sec:Spin}), and the total jet power. These estimates will require high-fidelity polarized and multi-frequency images with an angular resolution of ${\sim}15\,\mu{\rm as}$ (a spatial resolution of ${\approx}\,4GM/c^2$) and with sufficient dynamic range to simultaneously study both the near-horizon magnetosphere and the jet over many dynamical timescales.

\subsubsection{Physical Conditions and Launching Mechanisms for Relativistic Jets}
\label{sec:Jet_Conditions}

The ngEHT has the potential to substantially improve our understanding of the mechanisms that launch, collimate, and accelerate relativistic jets by measuring the physical conditions at the jet base. For instance, multi-frequency VLBI observations at cm-wavelength mainly probe the extended jet regions and have revealed that the energy distributions of relativistic electrons responsible for the emission follow power-laws. This is in marked contrast to the recent EHT observations of the horizon scale structure around \m87 and \sgra, which has been successfully modeled using thermal distributions of electrons \citep{EHTC_M87_V,EHTC_SgrA_V}. Important questions therefore arise regarding which physical mechanisms are able to accelerate the thermal particles, and where this particle energization occurs. Using multi-frequency observations at 86\,GHz and 230\,GHz while making use of VLBI synergies with the next-generation Very Large Array (ngVLA), the spectral index distribution of the radio emission can be mapped at high resolution, allowing estimates of the underlying eDF and indicating possible particle acceleration sites. In addition, linear polarization studies will reveal the magnetic field structure and strength in the jet, and circular polarization will reveal the plasma composition (leptonic/hadronic), opening a window to more detailed understanding of jet microphysics \citep[e.g.,][]{Emami_2023_plasma}. 

According to recent GRMHD models, a dynamic range of ${\sim}10^4$ will enable us to probe the jet in \m87 at a wavelength of 1.3\,mm on scales of hundreds of microarcseconds and to reliably measure the velocity profile. Besides the aforementioned array requirements, monitoring of the jet with cadences of days to weeks is required (for \m87, 1 day corresponds to roughly $3 t_{\rm g}$). \autoref{fig:Jet_Dynamics} shows simulated ngEHT reconstructions of the \m87 jet, illustrating the ability of the ngEHT to conclusively identify and track kinematic structure throughout the jet. Finally, in addition to \m87, there are several other potential AGN targets (e.g., Cen~A, 3C120, 3C84) of comparable BH mass and distance, which can also serve as laboratories to study jet launching activity. More distant AGN ($z > 0.1$) would require imaging on a ${\sim}$monthly basis.

\vspace{1cm}

%%%%%%%%%%%%%%%%%%%%%%%%%
\subsection{Transients}
\label{sec:Transients}
%%%%%%%%%%%%%%%%%%%%%%%%
Astrophysical transients are the sites of some of the most extreme physics in the present-day universe, including accreting sources such as BH X-ray binaries and Tidal Disruption Events, explosive events such as supernova as well as the LIGO/VIRGO gravitational wave bursts associated with neutron star-neutron star mergers such as GW170817 \citep{GW170817_a,GW170817_b}. 

In essentially all cases, the radio emission from these transients corresponds to synchrotron emission from relativistic electrons spiralling in magnetic fields either in a jet or in structures which have been energised by a jet associated with the transient \citep[e.g.,][]{Sari_1999,Mirabel_Rodriguez_1999,Fender_2006,Mooley_2018,Alexander_2020}. As with supermassive BHs in AGN, probing the formation, propagation and ultimate energetics of these jets is central to understanding the physics of BHs and how they convert gravitational potential energy of infalling matter into powerful collimated outflows. 

Because the field of astrophysical transients is so diverse, we have chosen to focus the ngEHT key science goals and associated requirement related to transients on two sets of objects: BH X-ray binaries (\autoref{sec:XRB}) and extragalactic transients (\autoref{sec:extragalactic_transients}). Together, these categories span most of the range both in the astrophysics under study and in the technical requirements for the ngEHT.

%%%%%%%%%%%%%%%%%%%%%%%%%%%%%%%%%%%%%%%%%%%%%%%%%%%%%%%%%
\subsubsection{Dynamics of Black Hole X-ray Binaries}
\label{sec:XRB}
%%%%%%%%%%%%%%%%%%%%%%%%%%%%%%%%%%%%%%%%%%%%%%%%%%%%%%%%%
Black hole X-ray binaries (BHXRBs) represent the bright end of the population of massive stellar remnants in our galaxy. They are expected to number in the few thousands among a likely population of ${\sim}10^8$ stellar mass BHs in our galaxy, with a typical mass around $7 M_\odot$. They accrete, usually intermittently, from a close binary companion and often reach accretion rates close to the Eddington limit. In other words, they are around five (eight) orders of magnitude less massive than \sgra (\m87) and are accreting at $>10^7$ times higher Eddington-ratioed rates. There are good reasons, and indeed much circumstantial evidence, to suggest that the coupling between accretion `states' and jet formation at high Eddington ratios are similar between supermassive and stellar-mass BHs, so their study genuinely, and dramatically, extends the parameter space of study of BHs \citep[e.g.,][]{Falcke_2004,Kording_2006}.

The event horizons of these stellar-mass BHs will likely never be resolvable by conventional telescopes, but remarkably it has been established that high-time resolution X-ray variability studies of BHXRBs probe the same range of scales in gravitational radii, $r_{\rm g} \equiv GM/c^2$, as the direct EHT imaging of \m87 and \sgra. Furthermore, decades of work has established good, but not yet precise enough, connections between characteristic patterns of variability, arising from within $100 r_{\rm g}$, and the formation and launch of the most powerful jets.

With the ngEHT, we will be able to probe BHXRB jets on scales around $10^6 r_{\rm g}$, at which scales bright \mbox{(sub-)mm} flares often have flux densities in excess of 1\,Jy and evolve considerably on timescales of \emph{minutes} \citep[e.g.,][]{Miller_Jones_2004,Tetarenko_2017,Tetarenko_2019}. VLBI studies of jets at 10 times larger angular scales have provided the most precise determination of jet launch time (and the corresponding activity in the accretion flow), evidence for strong directional variation and precession of the jet, and circumstantial evidence for interactions and---presumably---internal shocks between components moving at different speeds. This is also the region in $r_{\rm g}$ in which the jets of \m87 and other AGN have been seen to switch from an initially parabolic to a later conical cross section \citep[e.g.,][]{Asada_2012,Hada_2013,Tseng_2016,Kovalev_2020,Okino_2022}. With the ngEHT, we can directly test if this same collimation is occurring in BHXRB jets. Finally, we now know from the ThunderKAT project on MeerKAT \citep{Fender_2016} that large-scale jets from BHXRBs which decelerate and terminate in the ISM on timescales of ${\sim}1$\,year are common (rate of 2-4/yr): therefore only in this class of object can we track events from their creation and launch in the accretion flow through to their termination, providing an opportunity for precise calorimetry of their kinetic power.

\subsubsection{Extragalactic transients}
\label{sec:extragalactic_transients}

The broad term of extragalactic transients encompasses sources including Gamma Ray Bursts (GRBs), Tidal Disruption Events (TDEs), neutron star mergers, supernovae, fast radio bursts (FRBs), fast blue optical transients (FBOTs) and other related phenomena. The origin of the radio emission from these objects is often within relativistic jets, but it may also be more (quasi-)spherical.

Some of these phenomena remain optically thick and bright at (sub-)mm wavelengths for a considerable period of time \citep[months; e.g.,][]{Ho_2019,Margutti_2019} which places far less stringent requirements for response and scheduling of ngEHT. Nevertheless, there is a wide range of important physics which could conceivably be tackled, such as whether or not jets are being produced commonly (very topical for TDE jets, which may even be associated with neutrino production) and how much kinetic power was released in the event. Thus, the ngEHT could make significant discoveries by measuring the kinetic power, physical structure, and velocity in extragalactic transients such as GRBs, GW events, TDEs \citep[e.g.,][]{Curd_2022}, FRBs, and FBOTs

\subsection{New Horizons}
\label{sec:New_Horizons}

The ``New Horizons'' SWG was formed to explore and assess non-traditional avenues for ngEHT scientific breakthroughs. This group has examined topics including terrestrial applications such as planetary radar science, geodesy, and improved celestial reference frames \citep{Eubanks_2020,ICRC3}; studies of coherent sources including magnetars, masers, and fast radio bursts; and precise astrometry of AGN \citep{Reid_Honma_2014}. We now describe the two key science goals that have been identified by this SWG, both with cosmological applications: measurements of proper motion and parallax for a sample of AGN at distances up to ${\sim}80\,{\rm Mpc}$ (\autoref{sec:AGN_astrometry}), and studies of SMBHs and their accretion disks using water vapor megamasers, which can provide accurate measurements of the Hubble constant up to distances of ${\sim}50\,{\rm Mpc}$ (\autoref{sec:masers}).

\subsubsection{Proper Motions and Secular (CMB) parallaxes of AGN}
\label{sec:AGN_astrometry}

The multi-band capabilities of the ngEHT will enable the use of the source-frequency phase referencing \citep[SFPR;][]{Rioja_Dodson_2011} technique, potentially achieving ${\sim}\mu$as-level astrometry for targets that are sufficiently bright and close to known reference sources \citep{Jiang_2023,Rioja_2023}.  In addition to many other scientific applications such as measurements of (chromatic) AGN jet core shifts \citep[e.g.,][]{Sokolovsky_2011,Pushkarev_2012,Jiang_2021} and the (achromatic) orbital motions of binary SMBH systems \citep[e.g.,][]{DOrazio_2018}, one of the opportunities afforded by this astrometric precision is a measurement of the so-called ``cosmological proper motion'' \citep{Kardashev_1986} or ``secular extragalactic parallax'' \citep{Paine_2020}.  Because the Solar System is moving with respect to the cosmic microwave background (CMB) with a speed of $\sim$370\,km\,s$^{-1}$ \citep{Hinshaw_2009}, extragalactic objects in the local Universe 
should exhibit a contribution, $\mu_{\rm sec}$, to their proper motion from the Solar Systems's peculiar motion:  
\begin{equation}
    \mu_{\rm sec} \approx \left( 0.018 \text{\,$\mu$as\,yr$^{-1}$} \right) \left( \frac{H_0}{70 \text{\,km\,s\,Mpc$^{-1}$}} \right) \frac{|\sin(\beta)|}{z}\,,
\end{equation}
where $z$ is the object's cosmological redshift, $H_0$ is the Hubble constant, and $\beta$ is the angle between the location of the source and the direction of the Solar System's motion with respect to the CMB \citep{Kardashev_1986}. An object located at a distance of 10\,Mpc ($z \approx 0.0023$) is thus expected to have $\mu_{\rm sec} \sim 8\,\mu{\rm as}\,{\rm yr}^{-1}$, while an object located at a distance of 100\,Mpc ($z \approx 0.023$) is expected to have a proper motion of $\mu_{\rm sec} \sim 0.8\,\mu{\rm as}\,{\rm yr}^{-1}$.  By measuring the proper motion of many objects and using multi-frequency observations to mitigate chromatic effects in time-variable core shift effects \citep[see, e.g.,][]{Plavin_2019_coreshift}, the ngEHT could thus isolate the contribution of $\mu_{\rm sec}$ and provide coarse estimates of $H_0$ that are independent of standard methods \citep[e.g.,][]{Riess_2019,Planck_2020,Pesce_2020,Wong_2020}.

\begin{figure*}
    \centering
    \includegraphics[width=0.16\textwidth]{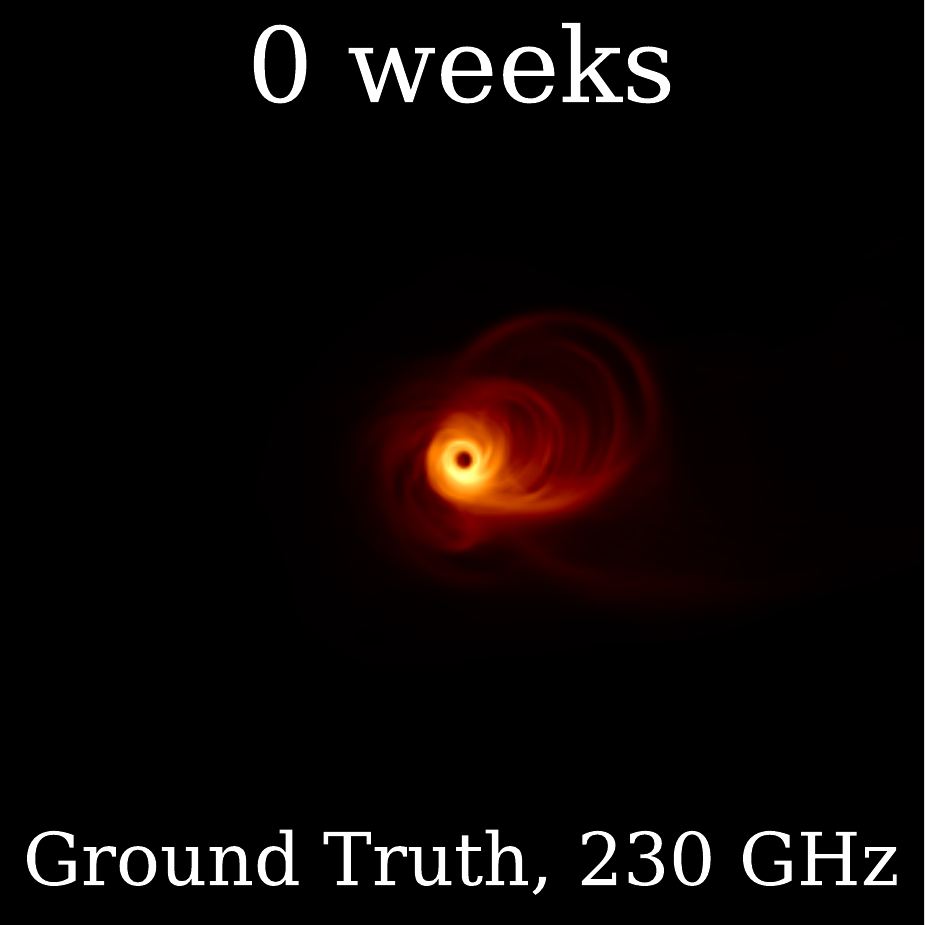}%
    \includegraphics[width=0.16\textwidth]{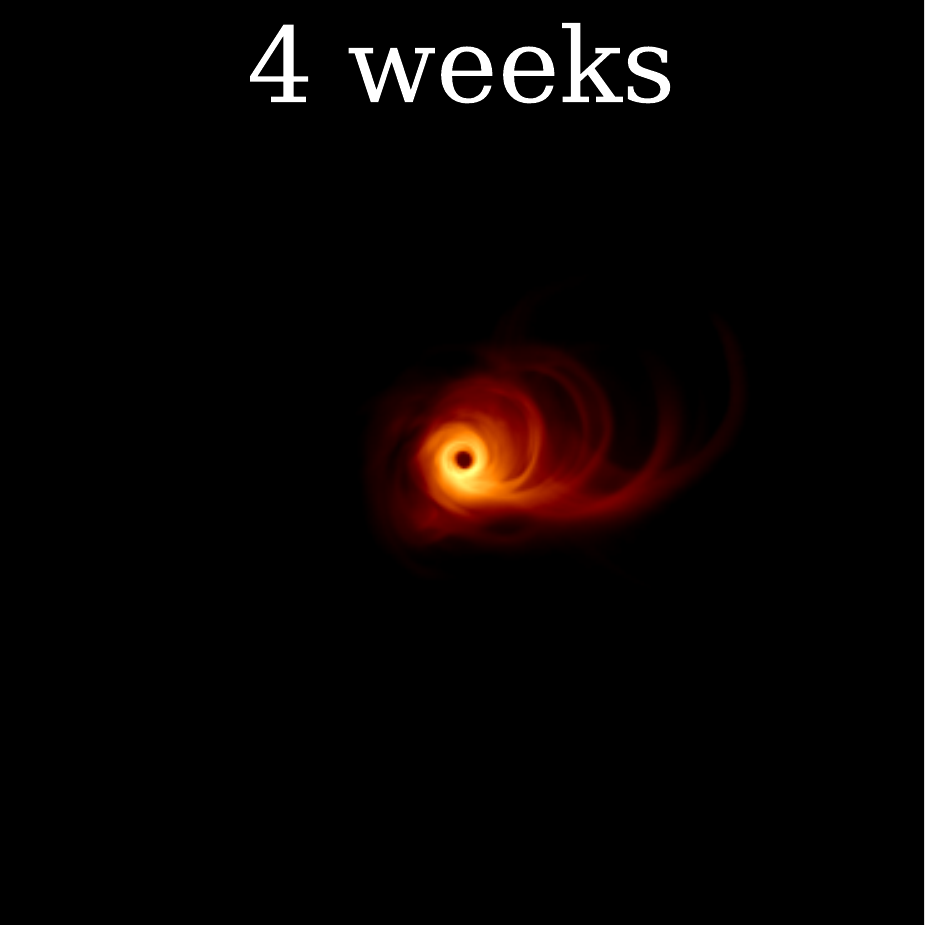}%
    \includegraphics[width=0.16\textwidth]{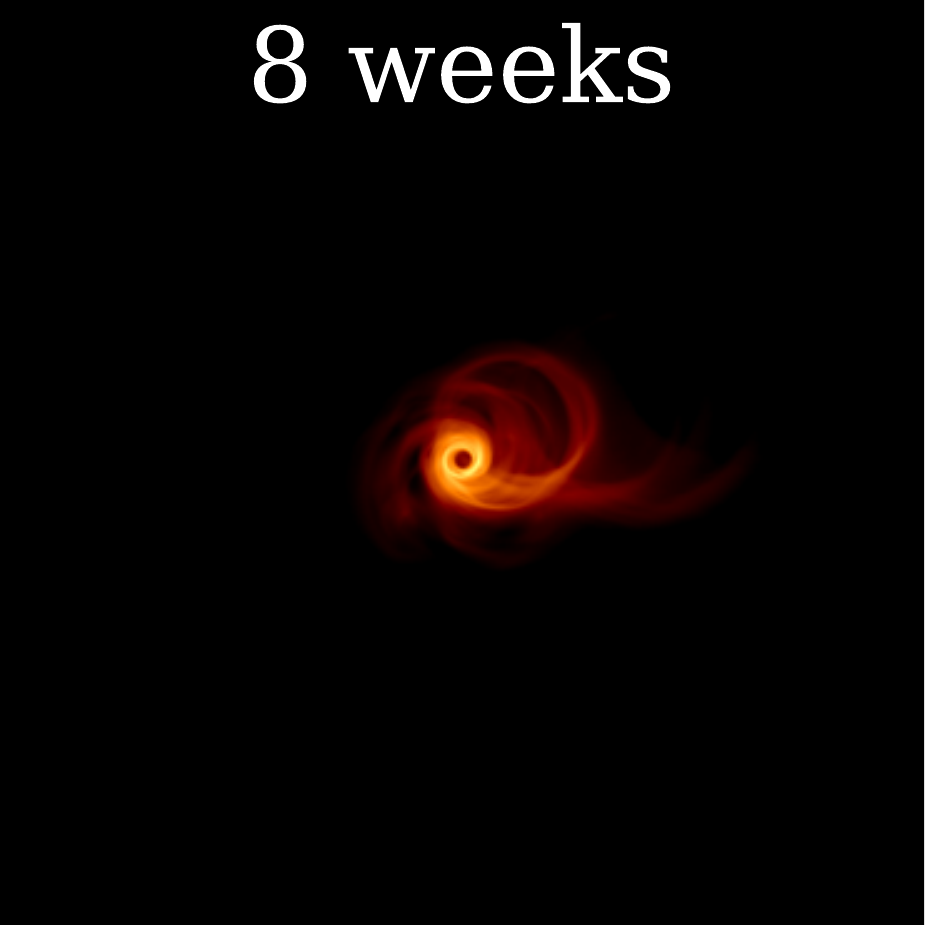}%
    \includegraphics[width=0.16\textwidth]{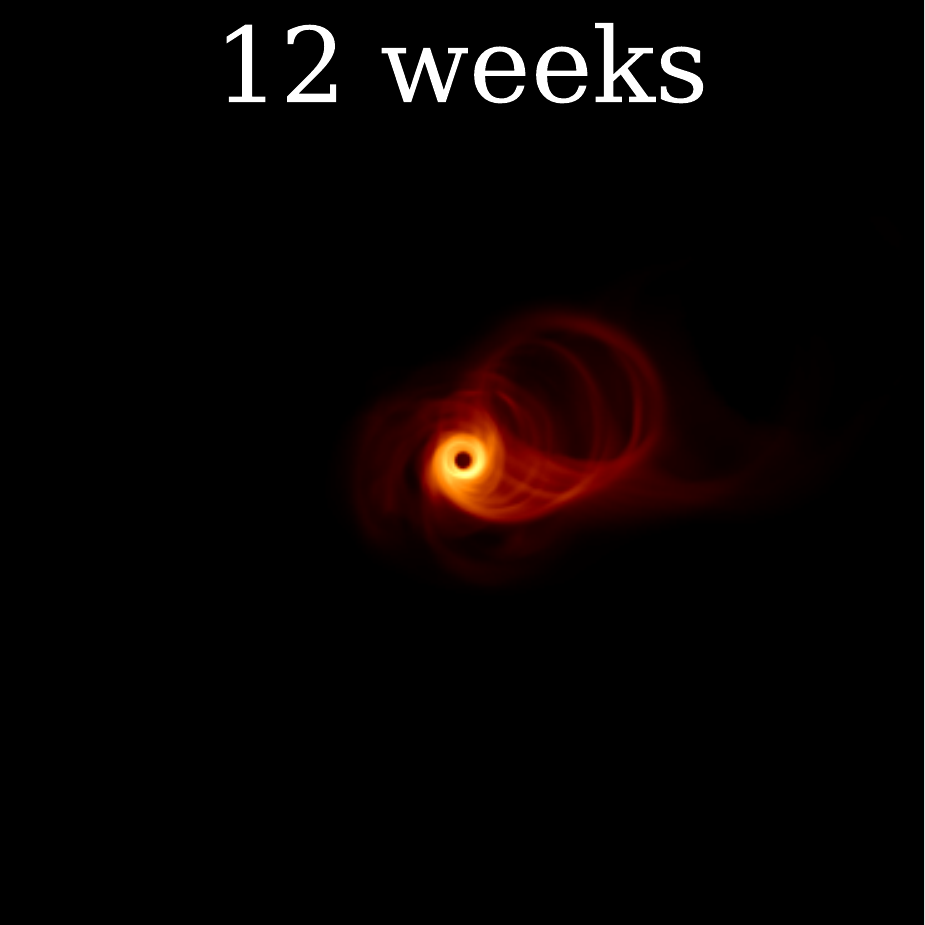}%
    \includegraphics[width=0.16\textwidth]{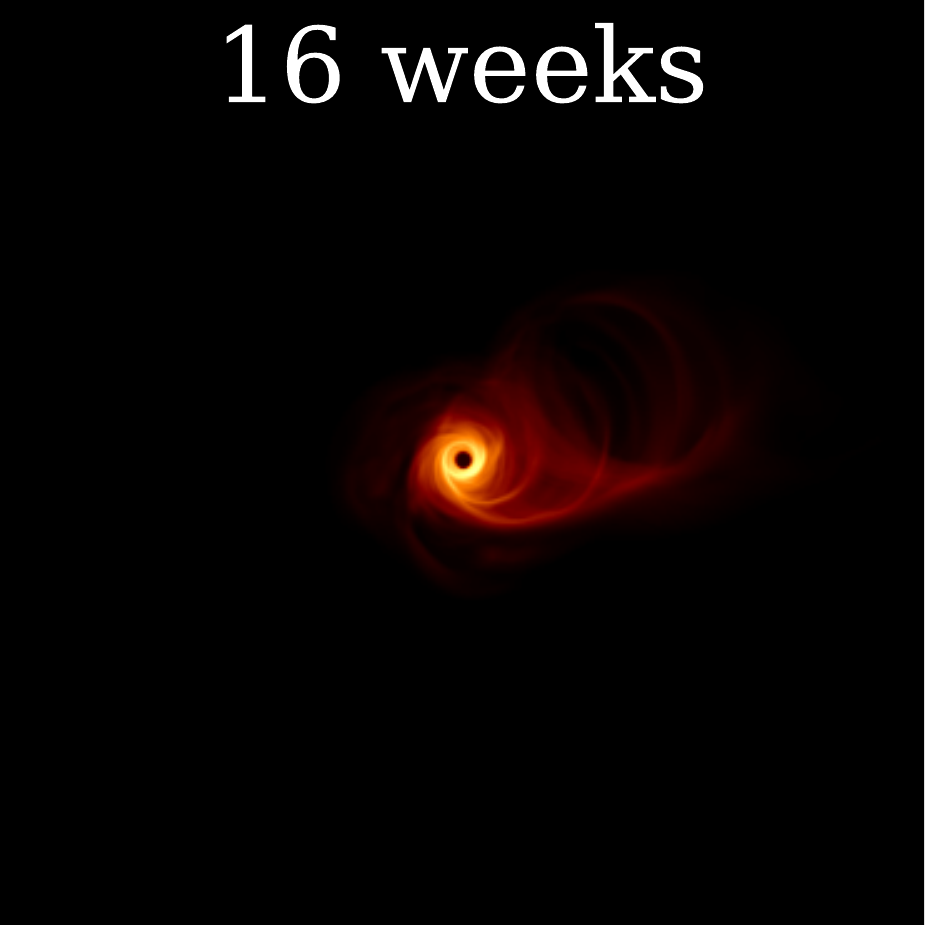}%
    \includegraphics[width=0.16\textwidth]{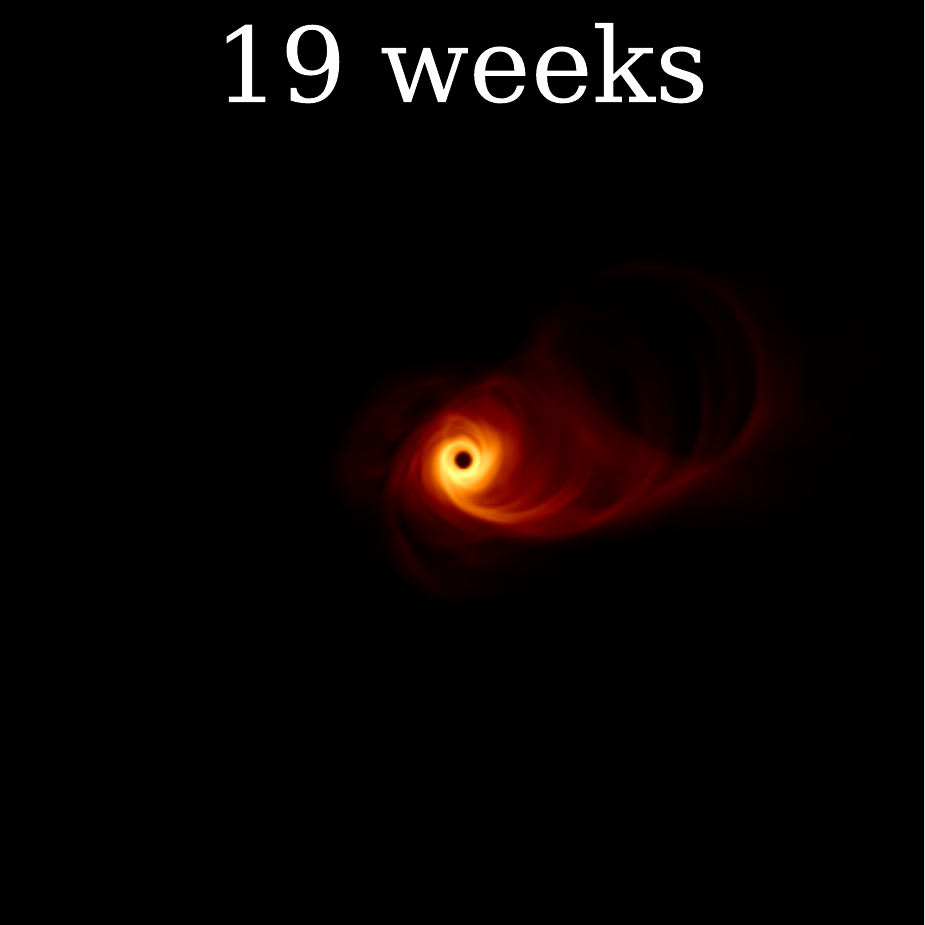} \\
    \includegraphics[width=0.16\textwidth]{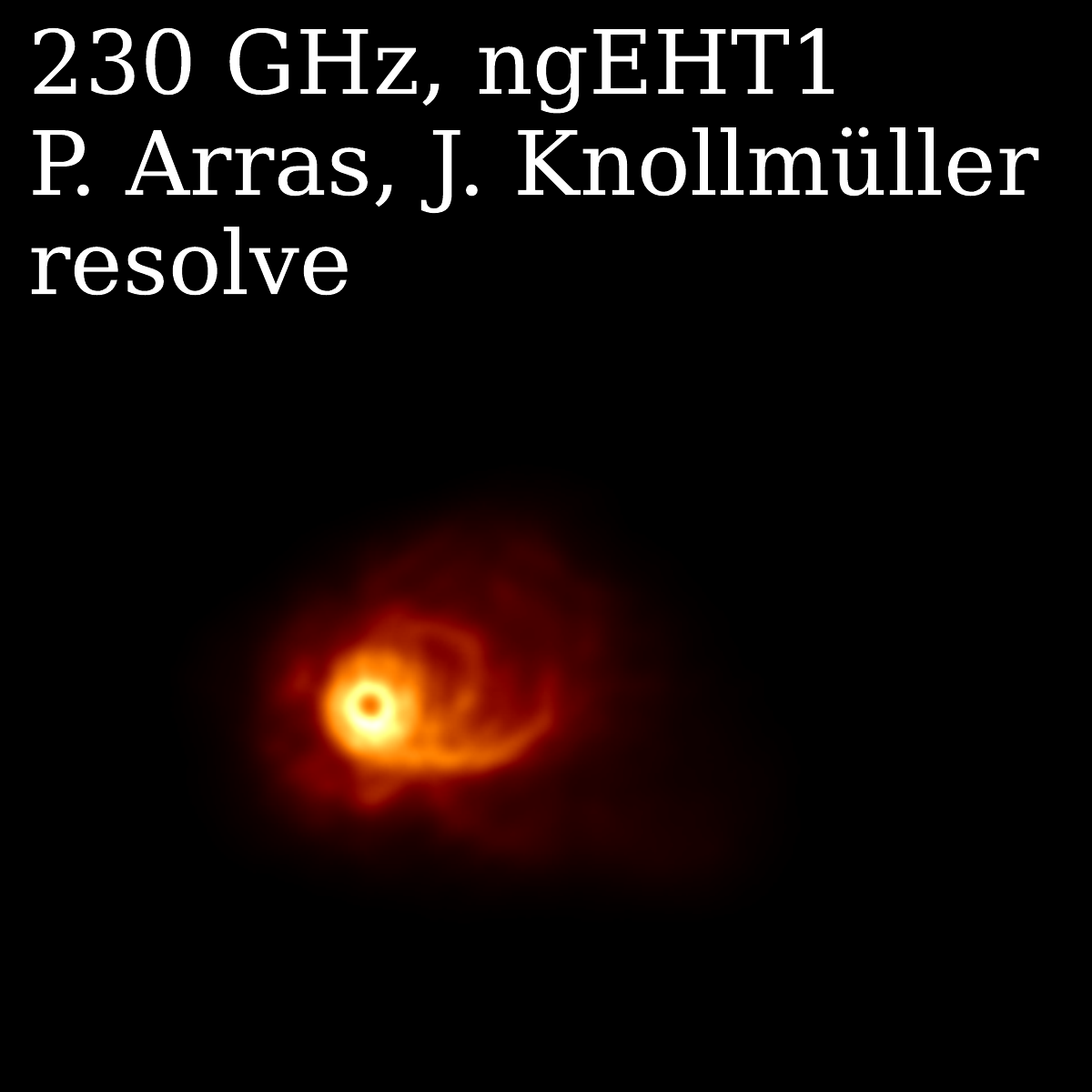}%
    \includegraphics[width=0.16\textwidth]{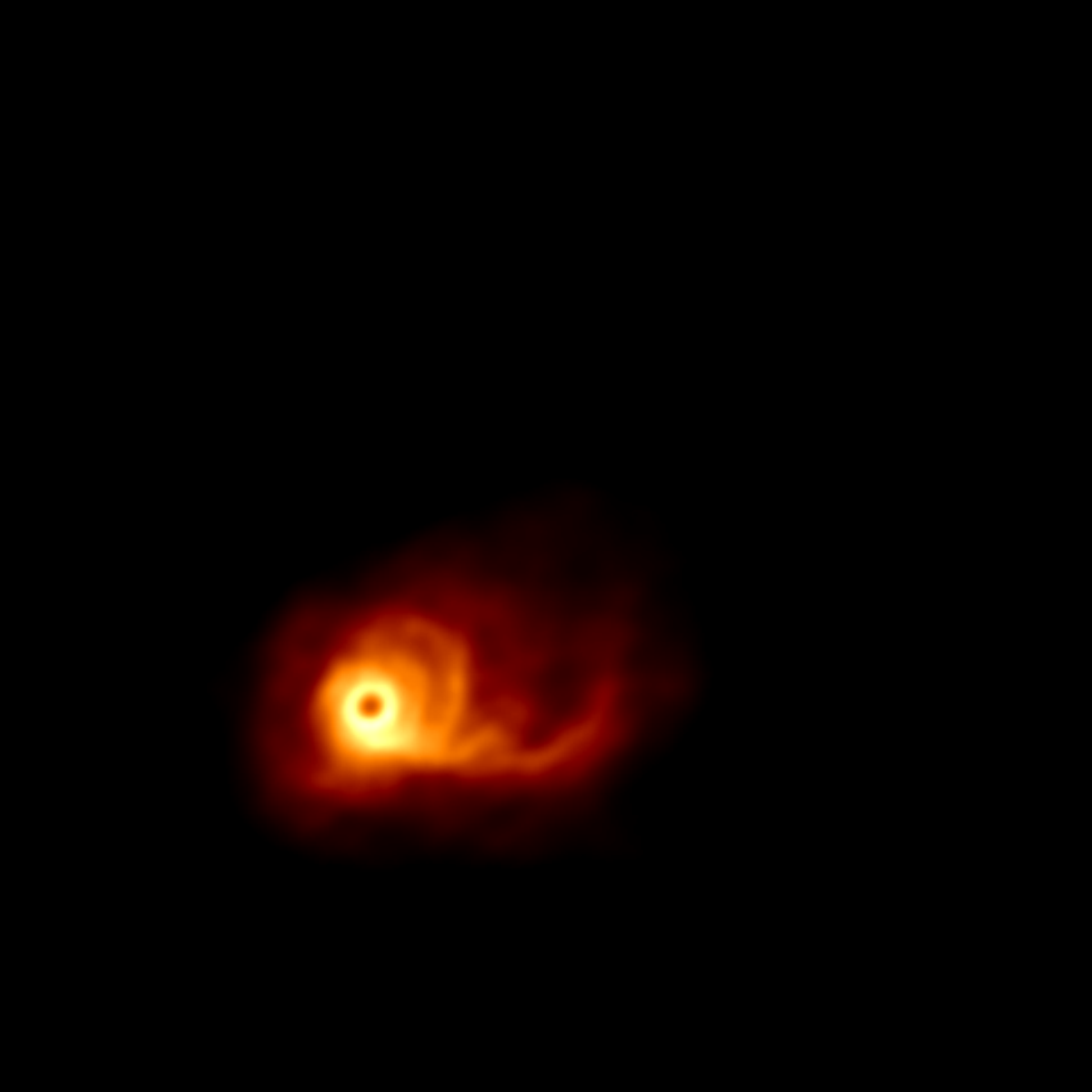}%
    \includegraphics[width=0.16\textwidth]{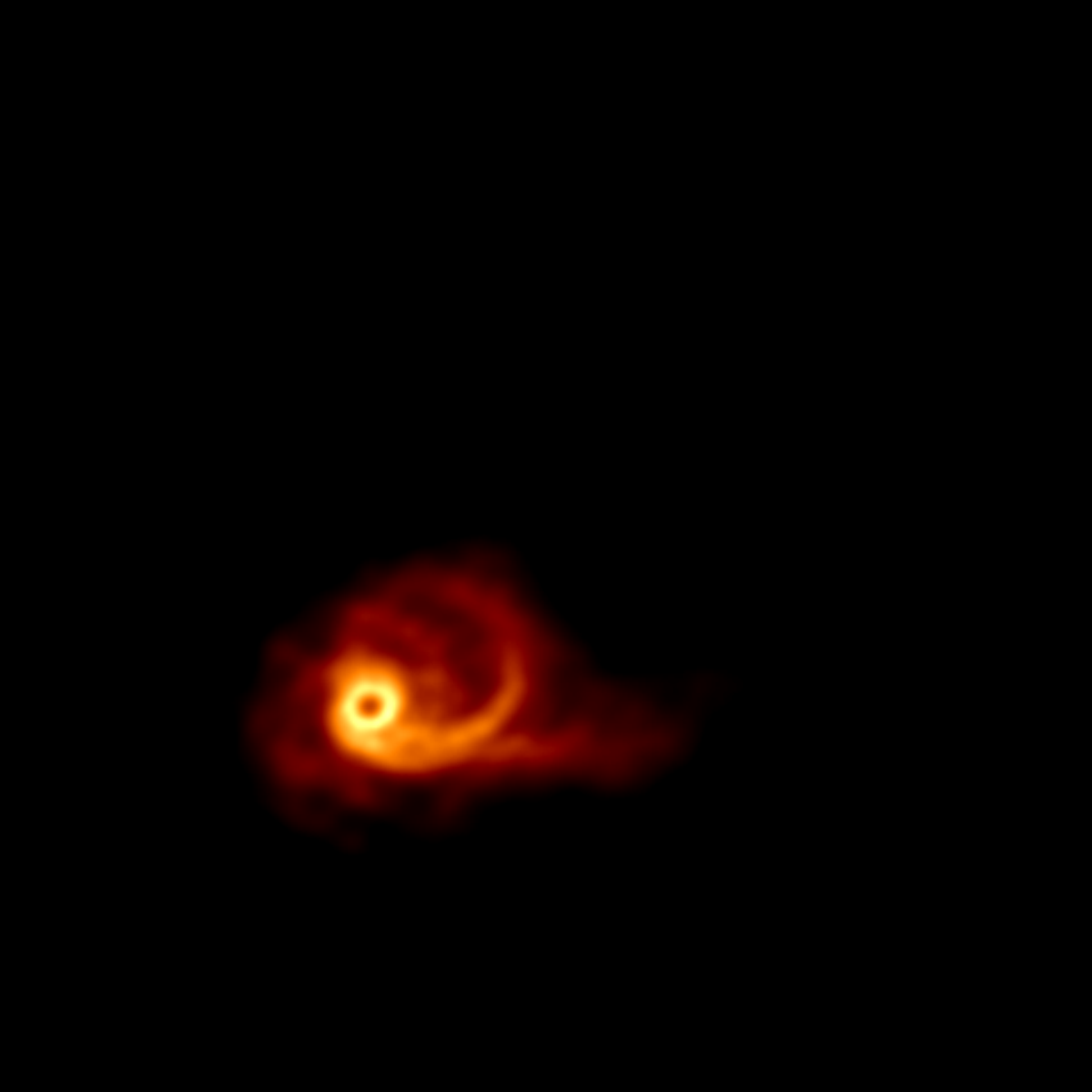}%
    \includegraphics[width=0.16\textwidth]{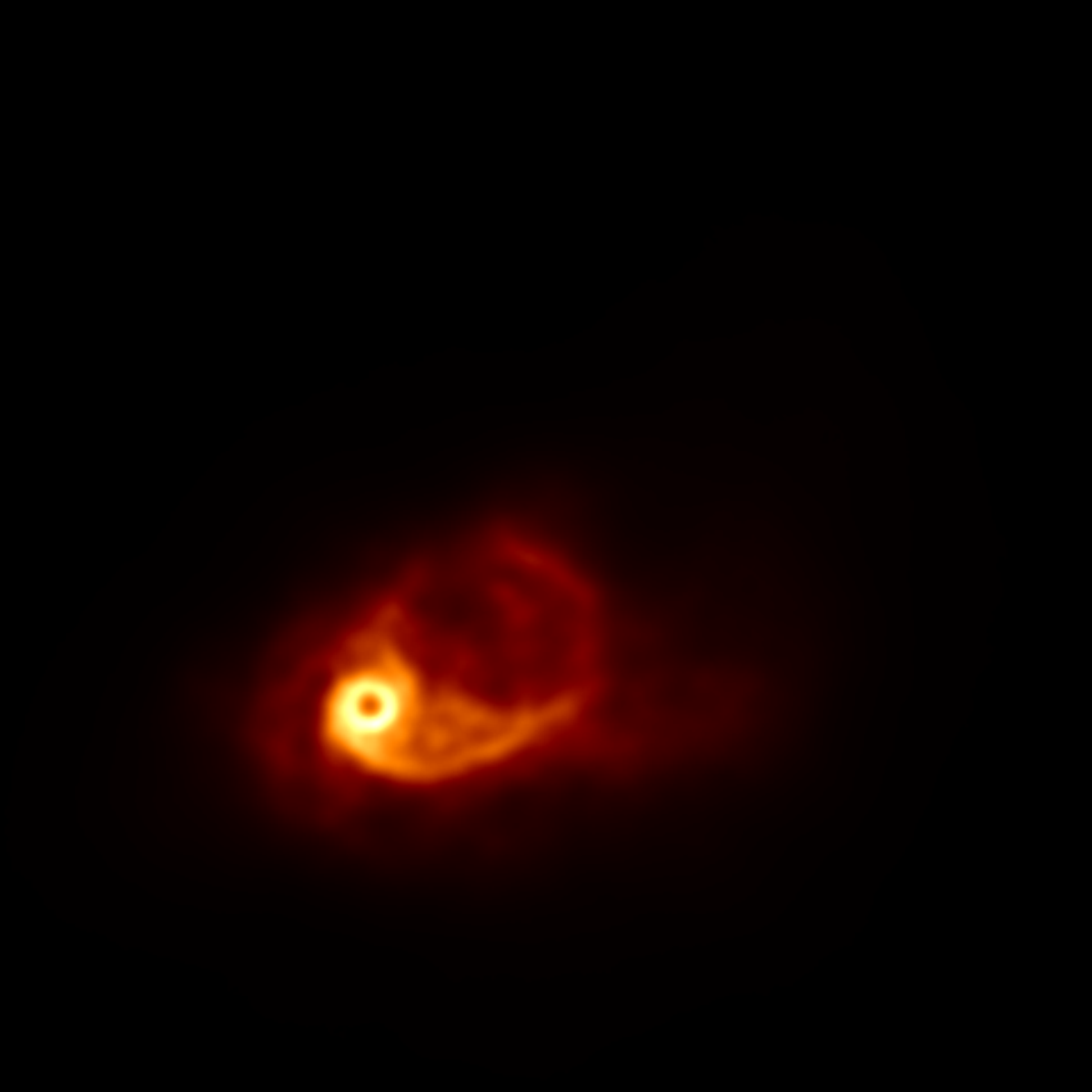}%
    \includegraphics[width=0.16\textwidth]{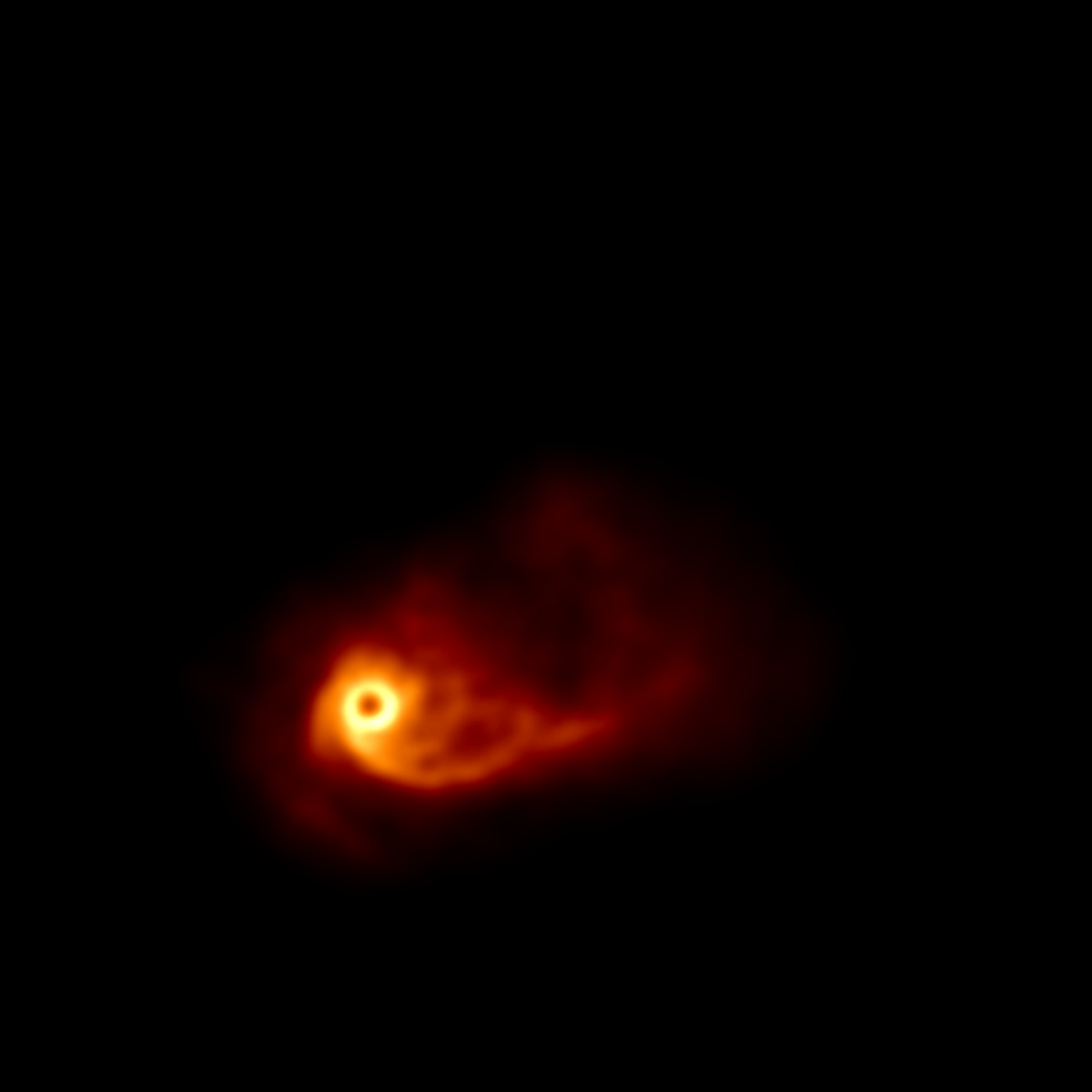}%
    \includegraphics[width=0.16\textwidth]{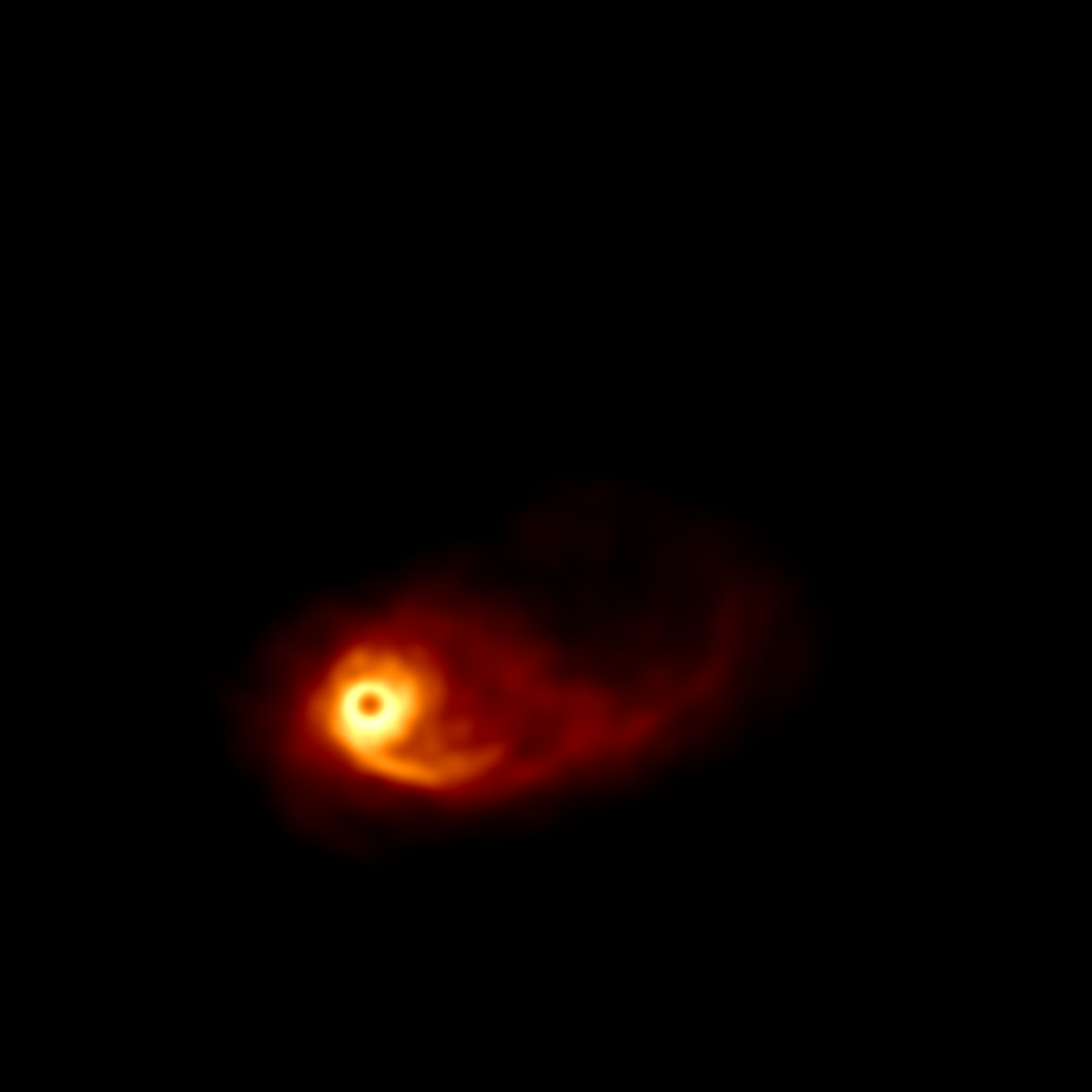} \\
    \vspace{4mm}
    \includegraphics[width=0.12\textwidth]{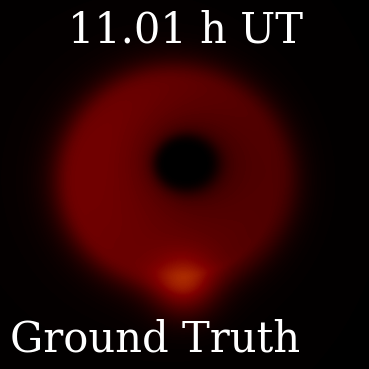}%
    \includegraphics[width=0.12\textwidth]{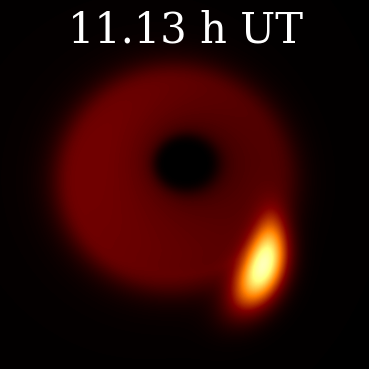}%
    \includegraphics[width=0.12\textwidth]{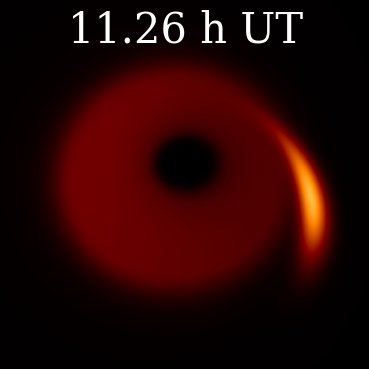}%
    \includegraphics[width=0.12\textwidth]{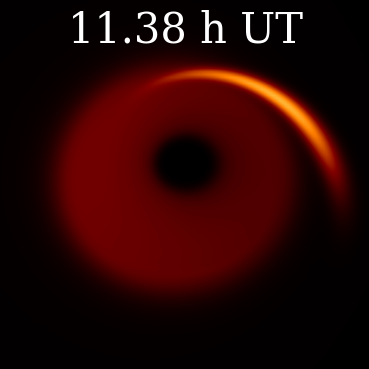}%
    \includegraphics[width=0.12\textwidth]{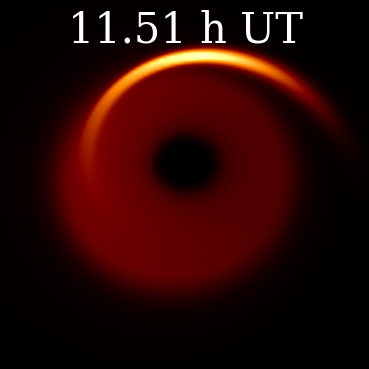}%
    \includegraphics[width=0.12\textwidth]{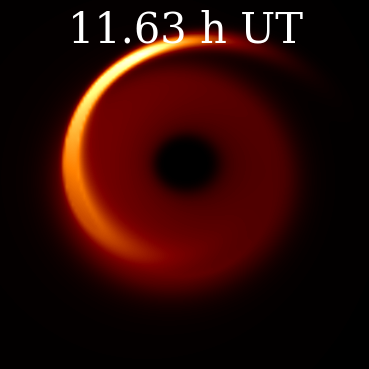}%
    \includegraphics[width=0.12\textwidth]{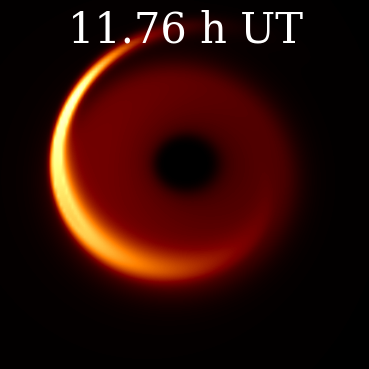}%
    \includegraphics[width=0.12\textwidth]{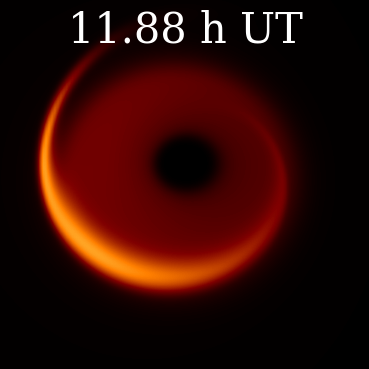} \\
    \includegraphics[width=0.12\textwidth]{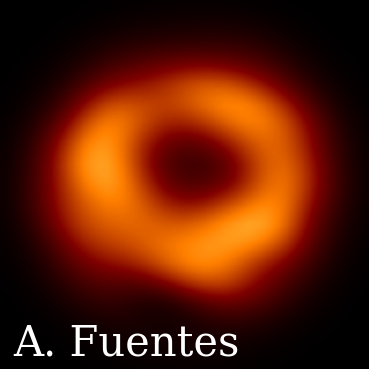}%
    \includegraphics[width=0.12\textwidth]{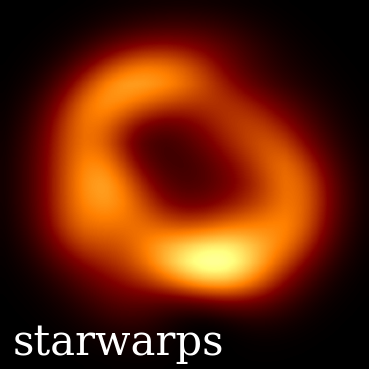}%
    \includegraphics[width=0.12\textwidth]{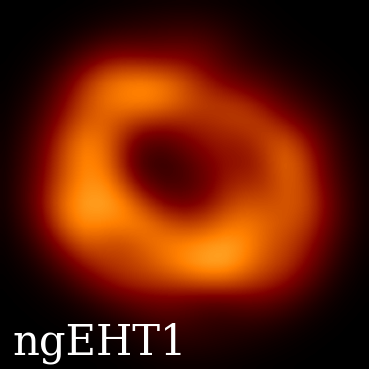}%
    \includegraphics[width=0.12\textwidth]{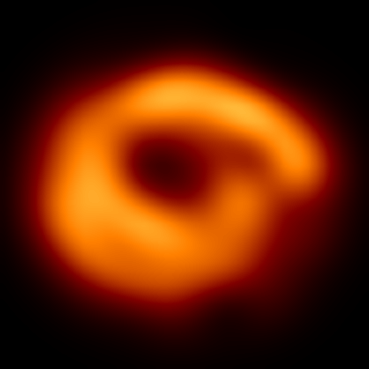}%
    \includegraphics[width=0.12\textwidth]{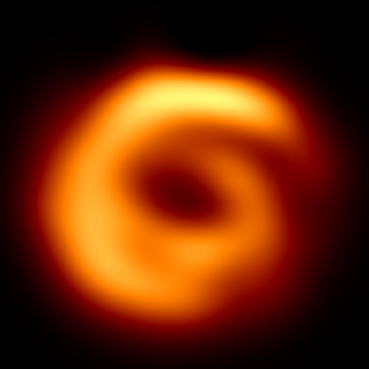}%
    \includegraphics[width=0.12\textwidth]{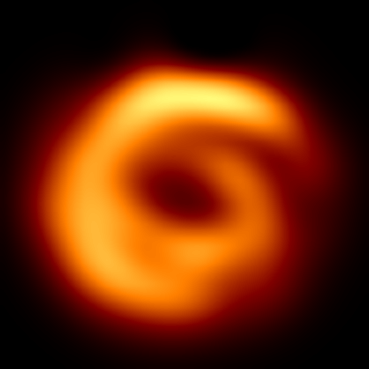}%
    \includegraphics[width=0.12\textwidth]{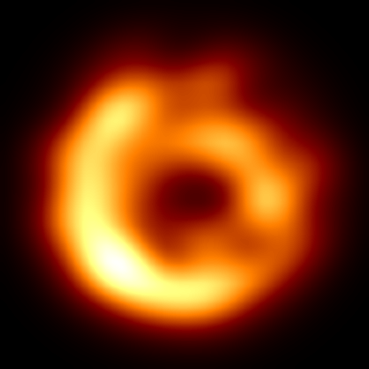}%
    \includegraphics[width=0.12\textwidth]{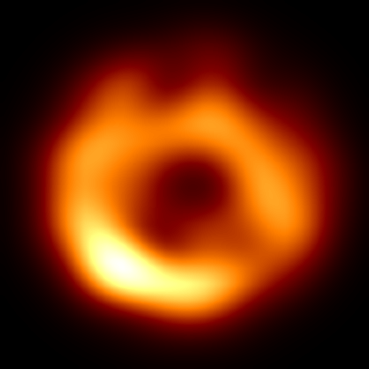}% \\
    \caption{Example ngEHT reconstructions for \sgra (top two rows) and \m87 (bottom two rows), using submissions for the second ngEHT Analysis Challenge \citep{Roelofs_2023}. For each source, upper panels show ground truth movie frames, and lower panels show example reconstructions. The \m87 ground truth movie is a GRMHD simulation generated with H-AMR \citep{Liska_2022} and ray-traced with {\tt ipole} \citep{Moscibrodzka_2018}; the reconstructed movie was produced using {\tt resolve} \citep{Arras_2022}. The \sgra simulation is a semi-analytic accretion flow with a shearing hot spot \citep{Broderick_2016, Tiede_2020}; the reconstructed movie was produced using {\tt StarWarps} \citep{Bouman_2018}. Panels are reproduced from \citet{Roelofs_2023}.}
    \label{fig:analysis_challenge}
\end{figure*}

\subsubsection{Studies of Black Hole Masses and Distances with Megamasers}
\label{sec:masers}

Water vapor megamasers residing in the molecular disks around nearby AGNs on scales of ${\sim}0.1\,{\rm pc}$ (${\sim}10^5\,r_{\rm g}$) 
have proven to be powerful tools for making precise measurements of SMBH masses \citep[e.g.,][]{Miyoshi_1995,Kuo_2011}, geometric distances to their host galaxies \citep[e.g.,][]{Herrnstein_1997,Braatz_2010}, and the Hubble constant \citep[e.g.,][]{Reid_2013,Pesce_2020}.  While the majority of research carried out to date has utilized the 22\,GHz rotational transition of the water molecule, other transitions are expected to exhibit maser activity under similar physical conditions as those that support 22\,GHz masers \citep{Yates_1997,Gray_2016}.  In particular, both the 183\,GHz \citep{Humphreys_2005,Humphreys_2016} and the 321\,GHz \citep{Hagiwara_2013,Pesce_2016,Hagiwara_2016,Hagiwara_2021} transitions have been observed as masers towards AGN.  The latter transition falls in the ngEHT observing band, as does another tranisition at 325\,GHz that is also expected to exhibit maser activity \citep{Kim_2023}.

Observations of water megamaser systems with the ngEHT will necessarily target transitions such as those at 321\,GHz and 325\,GHz, rather than the transition at 22\,GHz.  If the submillimeter systems are as bright as those at 22\,GHz, then the $>$order-of-magnitude improvement in angular resolution brought about by the ngEHT will impart a corresponding improvement in the precision of maser position measurements in these systems.  However, the typical brightness of submillimeter megamaser systems is currently unknown, and the two sources that have to date been observed at 321\,GHz both exhibit fainter emission at 321\,GHz than at 22\,GHz \citep{Hagiwara_2013,Pesce_2016}; it is thus possible that systematically fainter submillimeter transitions (relative to 22\,GHz) will offset the improvement in position measurement precision through reduced signal-to-noise ratios.  Nevertheless, even comparable measurement precisions for submillimeter transitions will provide a statistical improvement in the mass and distance constraints for systems observed in multiple transitions.  Furthermore, because the optimal physical conditions (e.g., gas temperature and density) for pumping maser activity differ between the different transitions, simultaneous measurements of multiple transitions in a single source may be used to provide constraints on those physical conditions \citep{Yates_1997,Gray_2016}.  It is also possible that future surveys will uncover populations of AGN that exhibit submillimeter maser activity but no 22\,GHz emission, thereby increasing the sample of sources for which the megamaser-based measurement techniques can be applied.

%%%%%%%%%%%%%%%%%%%%%%%%%%%%%%%%%%%%%
\subsection{Algorithms \& Inference}
\label{sec:AI}
%%%%%%%%%%%%%%%%%%%%%%%%%%%%%%%%%%%%%
The results produced by the EHT collaboration have been enabled by a suite of new calibration, imaging, and analysis softwares, many of which were custom-built to tackle the unique challenges associated with the sparsity and instrumental corruptions present in EHT data as well as with the rapid source evolution and scattering in \sgra \citep[e.g.,][]{Fish_2014,Lu_2016,Chael_2016,Johnson_2016,Akiyama_2017a,Akiyama_2017b,Johnson_2017,Chael_2018,Blackburn_2019,Janssen_2019,Broderick_2020_Themis,Broderick_hybrid,Sun_2020,Park_2021,Pesce_2021_DMC,Arras_2022,Sun_2022,Janssen_2022}.  Many of the difficulties that motivated imaging developments for the EHT are expected to be compounded in ngEHT observations, with a large increase in data volume (increased bandwidth, more stations, and faster observing cadence), dimensionality (multi-frequency and multi-epoch), and requisite imaging fidelity (larger reconstructible field of view and higher imaging dynamic range).  The next generation of algorithmic development is already underway, with new data processing \citep{Hoak_2022,Yu_2023_GPU}, imaging \citep{Muller_2022,Chael_2022,Tiede_2022comrade}, machine learning \citep{Qiu_2022}, and full spacetime \citep{Broderick_2020_Themis,Tiede_2020,Palumbo_2022,Qiu_2022,Levis_2022} methods being designed to address the challenges and opportunities associated with ngEHT data.

\begin{table*}
\centering
\normalsize
     \begin{tabular}{|l|}
     \hline
 {\bf Threshold Science Goals}\\
$\bullet~$Establish the existence and properties of black hole horizons\\
$\bullet~$Measure the spin of a SMBH\\
$\bullet~$Reveal Black Hole-Galaxy Formation, Growth and Coevolution\\
$\bullet~$Reveal how BHs accrete material using resolved movies on event horizon scales\\
$\bullet~$Observe localized heating and acceleration of relativistic electrons on astrophysical scales\\
$\bullet~$Determine whether jets are powered by energy extraction from rotating BHs\\
$\bullet~$Determine the physical conditions and launching mechanisms for relativistic jets\\
\hline
\hline
 {\bf Objective Science Goals}\\
$\bullet~$Constrain the properties of a BH's photon ring\\
$\bullet~$Constrain ultralight boson fields\\
$\bullet~$Determine how SMBHs merge through observations of sub-parsec binaries\\
$\bullet~$Connect SMBHs to high-energy and neutrino events within their jets\\ 
$\bullet~$Detect frame dragging within the ergosphere of a rotating BH\\
$\bullet~$Measure the inner jet structure and dynamics in BH X-ray binaries\\
$\bullet~$Detect the kinetic power, physical structure, and velocity in extragalactic transients\\
$\bullet~$Detect proper motions and secular (CMB) parallaxes of AGN up to $\sim$80 Mpc distances\\
$\bullet~$Leverage AGN accretion disk megamasers to measure their AGN host properties\\
\hline
\end{tabular}
\caption{{\bf Key Science Goals of the ngEHT}.}
 \label{tab:KSG}
 \end{table*}

\begin{table*}[t]
\begin{center}
\includegraphics[width=\textwidth]{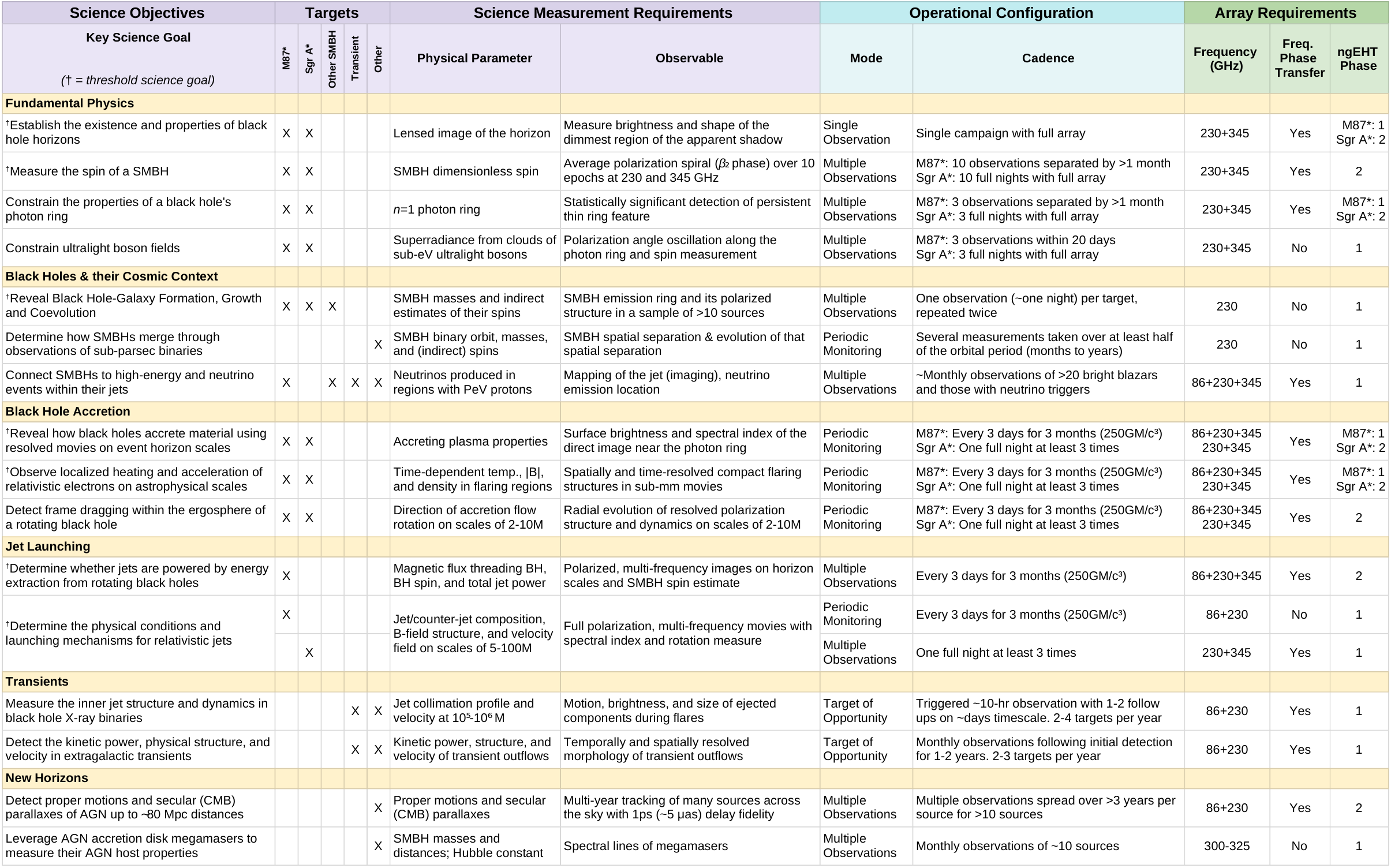}
\caption{
Representative subset of the ngEHT Science Traceability Matrix (STM). Daggers ($\dag$) indicate threshold science goals. 
The STM is used to guide the array design and to inform decisions about the multi-phase deployment.
}
\label{fig:STM}
\end{center}
\end{table*}

To assess scientific potential of the ngEHT, inform array design, and prompt the development of new algorithms, the ngEHT has launched a series of Analysis Challenges \citep{Roelofs_2023}. For each challenge, synthetic (ng)EHT datasets are generated from theoretical source models. These datasets are made available through the ngEHT Analysis Challenge website\footnote{\url{https://challenge.ngeht.org/}} and are accessible to anyone upon request. Participants then analyze the data by, e.g., reconstructing an image or fitting a model, and submit their results through the website. All submissions are evaluated with metrics quantifying, e.g., data fit quality or similarity of image reconstructions to the ground truth source model. 

Challenge~1 focused on static source models of \sgra and \m87 at 230 and 345\,GHz, and was set up mainly to test the challenge process and infrastructure. Challenge~2 was more science oriented, and focused on movie reconstructions from realistic synthetic observations of \sgra and \m87 at 86, 230, and 345\,GHz. Both challenges received submissions from a broad array of reconstruction methods. \autoref{fig:analysis_challenge} shows two submitted movie reconstructions from Challenge~2. The \m87 reconstruction shows the ngEHT's ability to reconstruct both the BH shadow and extended jet dynamics at high dynamic range, allowing detailed studies of jet launching. The \sgra shearing hotspot reconstruction, based on \citep{Tiede_2020} and motivated by the observational results of \citet{GRAVITY_2018b}, shows the ngEHT's ability to reconstruct rapid (intra-hour) accretion dynamics, even in moderate weather conditions at 230 GHz. In general, \citet{Roelofs_2023} found that standalone 345\,GHz imaging of the \m87 jet or \sgra dynamics is challenging due to severe atmospheric turbulence and optical depth effects. However, multi-frequency reconstructions showed that by utilizing information from 86 and 230\,GHz, the \m87 jet may be reconstructed at 345\,GHz \citep[see also][]{Chael_2022}. Also, while the \sgra shearing hotspot orbit could be reconstructed well, variability in GRMHD simulations was found to be more challenging to reconstruct due to the more turbulent nature of the plasma.

Two additional challenges are being run. Challenge~3 focuses on polarimetric movie reconstructions, and Challenge~4 will focus on science extraction, particularly to attempt measurements of the BH photon ring and the spacetime parameters. The merit of frequency phase transfer techniques for multi-frequency imaging will also be investigated \citep[see also][]{Issaoun_2023}.

\subsection{History, Philosophy, \& Culture}
\label{sec:HPC}

The History, Philosophy, \& Culture (HPC) SWG includes scholars from the humanities, social sciences, and sciences. HPC Key Science Goals were developed across four focus groups: Responsible Siting (\autoref{sec:siting}), Algorithms, Inference, and Visualization (\autoref{sec:aiv}), Foundations (\autoref{sec:foundations}), and Collaborations (\autoref{sec:collaborations}). We will now briefly summarize a selection of these goals that have been prioritized; for a more complete description, see \citetalias{HPC}.

\subsubsection{Responsible Siting}
\label{sec:siting}

Telescope siting has, historically, relied almost entirely upon ensuring that sites meet technical specifications required for observation including weather, atmospheric clarity, accessibility, and cost. As the issues at Mauna kea in Hawai’i show, \footnote{Two excellent doctoral dissertations offer fine-grained analysis of the mountaintop dispute, and are a good entry point into this issue. \cite{swanner2013mountains} focuses on the triply conflicting astronomical, environmental and indigenous narratives that collided at Mt. Graham, Mauna Kea, and Kitt Peak; \cite{salazar2014multicultural} addresses the Kanaka rights claim, specifically about the Thirty Meter Telescope (TMT), in opposition to a framing of the dispute as one of “stakeholders” or a “multicultural” ideal. \cite{swanner2017instruments} focuses on Mauna Kea in a subsequent article, also on the TMT. An important current Hawaiian-led impact assessment of the TMT, including further links, is \cite{kahanamoku2020native}; other Native Hawaiian scientists, including \cite{alegado2019telescope}  have spoken for a much-changed process and against the notion that opposition to the TMT is against science.} telescopes exist within a broader context and, as they choose sites, scientific collaborations incur the obligation to address ethical, social, and environmental specifications alongside technical ones. 

The ngEHT has already hosted a workshop dedicated to advancing responsible siting practices, which drew together experts in a wide range of fields including philosophy, history, sociology, advocacy, science, and engineering. \footnote{The workshop was held on the 4th of November 2022. Workshop Speakers included C. Prescod-Weinstein, K. Kamelamela, H. Nielson, M. Johnson, J. Havstad, T. Nichols, R. Chiaravalloti, S. Doeleman, G. Fitzpatrick, J. Houston, A. Oppenheimer, P. Galison, A. Thresher and P. Natarajan. Much of the work being done by the responsible siting group owes its genesis in the excellent contributions of the speakers and attendees of the workshop and we are grateful for their past and ongoing contributions.} This workshop was run by a dedicated siting focus group within the ngEHT HPC SWG, aimed at addressing the broader impacts of constructing and operating the chosen sites, with the goal of guiding short- and long-term siting decisions. Of particular interest to the group is consultation with areas outside of astronomy which also face questions of responsible siting including biotechnology\footnote{For a detailed discussion of siting and community guidelines for gene-drive technology, for example, see \cite{singh2019informed}}, archaeology and paleontology\footnote{There is much discussion within these fields of how we ought to think about community-led and non-extractive science. Good starting places for the literature include \cite{watkins2005through} and  \cite{supernant2014challenges}}, physics\footnote{An outstanding example of joint concern crossing environmental, cultural, epistemic, and technical concerns, in the case of LIGO, can be found in \cite{Nichols2023Hidden}.  Another instanced of community participation by (here in relation to NASA for their Asteroid Redirect Mission): \cite{tomblin2017integrating}.  On the siting of the Superconducting Supercollider, \cite{riordan2015tunnel}; an historical-anthropological study of the placement of the French/European launch center, \cite{redfield2000space}}, and nuclear technologies\footnote{Consent, and environmental justice, have been at the center of siting nuclear facilities, including power generation, weapons testing, accident sites, and waste disposal. The literature is vast, but a starting point with many further references can be found in \cite{gerrard1996whose} an environmental lawyer addresses community concerns about siting;  \cite{Kuletz1998tainted} focuses on Western US nuclear sites of waste; \cite{masco2013nuclear} attends to the quadruple intersection of weapons scientists, Pueblo Indian nations, nuevomexicano communities, and activists as they live amidst and confront the legacy of Los Alamos. On consent-based siting rather than top-down imposition, see \cite{hamilton2012blue}; and for a recent development and analysis of consent-based siting, \cite{richter2022process}.}. Ultimately, the goal is to model the decision-making process by joining technical, environmental, and community concerns, and to arrive at explicit guidelines that could assist with future siting challenges.
 
 For the ngEHT to achieve its goals on responsible siting, a number of concrete steps are necessary. First, the collaboration must integrate social and environmental impacts into its siting decisions, initially, via the inclusion of ethicists, social scientists, environmental experts, and local community advocates in siting meetings who will contribute to the decision-making process as well as the inclusion of explicit cultural, social and environmental factors in siting decision metrics; later, via the creation and performance of explicit community impact studies, in addition to reviewing the environmental impact studies already performed as part of the standard siting. These studies will embrace surveys of local social factors for sites to aid in the decision process and will involve on-site community consultation as well as work with local government and academic structures. 
 
Second, the collaboration must ensure that when telescopes are built, they are done so in collaborative and non-extractive ways that are sensitive to the history and culture of local communities and the lands in consideration. This goal will require establishing an ongoing dialogue with local community groups as early as possible in the siting process, and setting up explicit agreements that are mutually beneficial to all parties. As such, funding for community consultation and projects is a central part of the funding structure for the ngEHT; the aim is to ensure that local educational, scientific, and economic opportunities are built into the project from the out-set. This will involve examining local relationships with existing sites to be supplemented with new technology, as well as forging new relationships where un-developed sites are under consideration. The ngEHT project will be carefully considering who is at the table, and ensuring all local groups that may be impacted have a voice in the process. The ngEHT will also aim to work to integrate local and traditional knowledge into this process, recognizing that these are not in tension with scientific knowledge, but are continuous with it. Moreover, each site will be unique, with different needs and histories that will inform the kinds of relationships that will develop. As such, part of the community impact study will need to detail what sort of benefits local communities may want from, as well as offer to, the ngEHT collaboration. Possibilities include improved infrastructure, education funding, outreach, and knowledge exchange done under terms and conditions that meet the needs of the communities in question.

The ngEHT must also accept the fact that community, environmental, and cultural aspects may prevent a site from being developed, and that a ‘no’ from locals is a legitimate outcome. A clear goal, then, is to work with community siting experts from both inside and outside astronomy to establish what a ‘no’ looks like, as well as a ‘yes’, and to develop norms and practices which can help survey local groups to ensure their voices are being heard. 

Third, the ngEHT aims to minimize its environmental impact, including careful consideration of how construction and development of sites may impact native ecosystems as well as actively planning for what the eventual decommissioning and subsequent environmental repair of a site will look like. The ngEHT is committed, wherever possible, to using environmentally friendly techniques, technology, and materials, including in energy-efficient data-storage and computing. 

Finally, a major goal of this focus group will be the production of one or more papers detailing current best-practices for responsible telescope siting. Here the initial three to five sites (i.e., those in the ngEHT Phase-1) will be treated as proof-of concept sites where norms can be designed and established, and experts from both inside and outside astronomy will be brought in to help guide the paper writing process.

\subsubsection{Algorithms, Inference, and Visualization}
\label{sec:aiv}

The ngEHT is a long-term project which will heavily rely on software supported modes of reasoning, including imaging algorithms for image reconstruction, and GRMHD simulations and relativistic ray tracing codes for parameter extraction. Philosophers of other sciences relying on computer simulations (including climate sciences and theoretical cosmology) diagnosed that problematic features might arise in such situations. These include \citep{frigg2015philosophy,winsberg2018philosophy} (i) kludging: temporary and ad hoc choices (concerning, e.g., values of parameters, or a manner of merging together two pre-existing fragments of code) made for convenience and without principled justification; (ii) generative entrenchment: contingent choices made during code development in order to deal with problems arising in particular contexts are baked in and transferred to future versions; over time, awareness of the origin of various fragments might be lost; (iii) confirmation holism: assigning success or failure of a numerical model as a whole to a particular fragment of code becomes very hard. Some of these problematic features may have positive elements --- for example, feature (i) makes code development faster than if it were properly documented; feature (ii) might represent consensus of the collaboration. Awareness of these features and development of active means of preventing their negative effects will make inference methods of the ngEHT more reliable.

Further, new inferential methods based on various forms of machine learning and artificial intelligence are becoming increasingly widespread, including in astronomy. Such methods come with many benefits, including much faster data processing times, but also with drawbacks, including lack of epistemic transparency (inner workings of a machine learning model are not easily available or even understood by its users --- in contrast with the steps taken by a more traditional imaging or parameter extraction algorithm), and risk of building in bias through training on data sets containing untested assumption about the target system. Frameworks for mitigating these risks, so-called explainable artificial intelligence, have been developed \citep[e.g.,][]{samek2019explainable,zednik2021solving,beisbart2022philosophy}. We will systematically evaluate these methods and motivations behind them, isolating those which can and which should be applied to future ngEHT data analysis pipelines.

Reception of astronomical images takes place in a broader context of visual culture, and we will consider the importance of aesthetic choices made during production, such as assignment of color to underlying physical parameters or landscape associations invoked by the resulting image (e.g., \citealt{kessler2012picturing}; \citetalias{HPC}). As for the EHT, images produced by the ngEHT will shape public perception of black holes and astronomy. Analysis of such cultural factors will help with being intentional about the impact and perception of images—inside and outside the technical community. Accordingly, procedures for systematically including such choices and for testing whether an image succeeds in conveying the intended connotations will be developed, and applied, for example, to future polarization data and multi-frequency images.

The requirement to achieve long-term reliability of ngEHT inferences will necessitate the identification of inference methods deemed undesirable, and development of software evaluation tests to ameliorate those features.  Improving image presentation will focus attention on cultural factors that shape audience reactions to visualizations—in turn, we will need to develop comprehension tests probing audience responses.

\subsubsection{Foundations}
\label{sec:foundations}
The Foundations focus group complements the Fundamental Physics working group, providing a different, critical lens for thinking about what the ngEHT observations can tell us about fundamental physics. The ngEHT results will both be informed by, and inform, philosophical and historical perspectives on issues such as scientific representation and modeling, idealization, underdetermination, theory testing (confirmation), and more \citepalias[\S 3]{HPC}. This focus group facilitates ongoing interdisciplinary discussions of foundational issues, in parallel with discussions in fundamental physics. 

\vspace{2em}

\subsubsection{Collaborations} 
\label{sec:collaborations}

A fully integrated working group of scholars from the social sciences and humanities within a STEM collaboration provides an unprecedented opportunity to optimize the collaboration structure from the very beginning. Our main goal is a structure that enables, encourages, and emphasizes transparent decision-making, diversity, fair credit assignment and accountability \citepalias[\S 4]{HPC}.\footnote{For lessons learnt regarding knowledge formation, governance, organisational structure, decision-making, diversity, accountability, creativity, credit assignment and the role of consensus, from a range of perspectives across the humanities and social sciences, see e.g.\ a) in general: \citet{Galison1992,KnorrCetina1999,Sullivan2009,Shrum2001,Boyer-Kassem2017} and references therein; b) for specific collaborations and institutions: \citet{Collins2017,Nichols2022} on LIGO; \citet{Boisot2011,Ritson2021,Sorgner2022,Merz2022} on ATLAS and/or CERN; \citet{Jebeile2020} on the IPCC; \citet{Smith1993,Vertesi2020} on NASA; and \citet{Traweek1988} on SLAC and KEK.} This translates directly into various requirements for the ngEHT collaboration, as detailed further below. 

In addition, a long-term Forecasting Tournament \citep{mellers2015identifying,camerer2018evaluating} will clarify the ngEHT decision-making process. Participants' judgments about the outcome of ngEHT experiments and observations will reveal the novelty of eventual results and will elucidate the process of hypothesis generation and testing. By systematically collecting predictions, we will be able to track the return on testing different hypotheses, identify unresolved ambiguities within the design or implementation of an experiment (which may lead to new areas of investigation) and develop a more (cost) effective research management strategy \citep{dellavigna2019predict}. There are also direct epistemic advantages to surveying predictions and expectations. The EHT already went to great lengths to counteract a quite natural tendency to halt image reconstruction when the images coincided with anticipated results \citepalias[for example, blind trials with known, simulated data; autonomous imaging groups who did not share intermediate results;][]{EHTC_M87_IV}. In the ngEHT, imaging programs will become more elaborate, use of AI more extensive, and data volumes will expand rapidly. Therefore it will be increasingly important for the collaboration to be aware of forecasted results---precisely to avoid premature confirmation. 

These goals require frequent monitoring and evaluation of the internal communication structure and climate. This will be achieved via the complementary methods of surveys, interviews, and network analysis tools from the digital humanities (\citetalias[\S 4.3]{HPC}).\footnote{Regarding network analysis, communication structures and epistemic communities, see for instance the following texts and references therein: \citet{Kitcher1990,Kitcher1993,zollman2007,zollman2010,Zollman2013,Longino2019,lalli_dynamics_2020,lalli_socio-epistemic_2020,light2021,wuethrich2022,Seselja2022}.} %I don't understand why this footnote doesn't justify/align correctly on the right.
This will require that at least some collaboration members be available for interviews and surveys. Only against the backdrop of this ongoing feedback loop will it be possible to ensure the long-term effectiveness of the following further requirements: a) a governance structure that includes a central, representative, elected body, as well as a standing ethics committee responsible for the creation, adherence to, and updating of the collaboration's Community Principles and Code of Conduct---see \citetalias[\S 4.4]{HPC} for a tentative proposal; b) an authorship and membership model tailored to the needs of a modern collaboration involving members from a diverse group of (scientific and non-scientific) cultures, i.e.\ a model that accounts for fair distribution of credit and accountability and allows for and realizes the value of dissenting opinions.\footnote{Regarding authorship challenges and possible solutions relevant to the ngEHT context, see e.g.\ \citet{Resnik1997,Rennie1997,Cronin2001,Galison2003,Wray2006,Boyer-Kassem2017,Mcnutt2018,Bright2018,Heesen2018,Dangphd,Nogrady2023,Habgood-Cooteforth} and \url{www.icmje.org/icmje-recommendations.pdf}.} A dedicated task force has begun developing such a model.

\section{Summary}
\label{sec:Summary}

The ngEHT project has undergone a multi-year design process to define community-driven science priorities for the array. This process has identified breakthrough science related to studies of BH spacetimes, as well as a wealth of new opportunities beyond what has been explored with past EHT experiments. These science opportunities arise from the potential to substantially expand upon the currently explored parameter space: 
\begin{itemize}
\item Improved angular resolution and image fidelity through increased sensitivity and baseline coverage. These enhancements are the most significant requirements for studies of fundamental physics with the ngEHT. 
\item Expanding from independent multi-band observations to simultaneous multi-band observations at 86, 230, and 345\,GHz. This upgrade will substantially improve the EHT's sensitivity to observe faint sources, dim extended emission, and compact structure on the longest baselines at 345\,GHz, especially through the use of multi-frequency phase transfer.
\item Adding more sites to enable ``snapshot'' imaging of variable sources including \sgra, and extending observing campaigns over multiple years. Together, these upgrades will improve the temporal sensitivity of current EHT observations by ${\sim}5$ orders of magnitude, enabling a wealth of new variability studies (see \autoref{fig:resolution_timescale}).
\end{itemize}

We have classified each of the key science goals discussed in \autoref{sec:Science_Drivers} as either {\it Threshold} or {\it Objective}. Threshold science goals define the minimum target that the array concept is designed to meet. Objective science goals are additional major science opportunities or stretch target for the array concept to meet. This classification does not indicate the relative merit of the science objective; some goals are assigned as objective because they are considered to be too speculative or high-risk (e.g., studies of the photon ring and frame dragging), insufficiently unique to the ngEHT (e.g., studies of axions and SMBH binaries), or too poorly understood to define a precise associated instrument requirement that will guarantee success (e.g., studies of extragalactic transients). \autoref{tab:KSG} provides the categorization of each goal. In addition, we have developed a set of homogeneous array requirements for the science goals in the framework of a Science Traceability Matrix (STM). A representative subset of the STM is given in \autoref{fig:STM}.

In conclusion, the ngEHT scientific community has identified a series of science objectives, with associated observational advances that are feasible over the coming decade. Taken together, they offer a remarkable opportunity to push the frontiers of VLBI and to enable a series of new discoveries that will elucidate the extraordinary role of BHs across all astrophysical scales.